\documentclass[twoside,twocolumn,9pt]{article}
\usepackage{extsizes}
\usepackage[super,sort&compress,comma]{natbib} 
\usepackage[version=3]{mhchem}
\usepackage[left=1.5cm, right=1.5cm, top=1.785cm, bottom=2.0cm]{geometry}
\usepackage{balance}
\usepackage{mathptmx}
\usepackage{authblk} 
\usepackage{sectsty}
\usepackage{graphicx} 
\usepackage{lastpage}
\usepackage{amsmath}
\usepackage{bm} 
\usepackage[format=plain,justification=justified,singlelinecheck=false,font={stretch=1.125,small,sf},labelfont=bf,labelsep=space]{caption}
\usepackage{float}
\usepackage{fancyhdr}
\usepackage{fnpos}
\usepackage[english]{babel}
\addto{\captionsenglish}{%
  
}
\usepackage{array}
\usepackage{droidsans}
\usepackage{charter}
\usepackage[T1]{fontenc}
\usepackage[usenames,dvipsnames]{xcolor}
\usepackage{setspace}
\usepackage[compact]{titlesec}
\usepackage{hyperref}

\usepackage{epstopdf}

\definecolor{cream}{RGB}{222,217,201}

\newcommand{\INTPHI}{\int\limits_0^{2\pi} d\phi}

\title{\LARGE\bfseries\noindent
\textit{E.\ coli} bacterium near corrugated surfaces: near-surface swimming, 
escape, and hydrodynamic trapping}

\author[1]{Pierre Martin}
\author[1]{Gon\c{c}alo C. Antunes }
\author[1]{Holger Stark}

\affil[1]{Institute of Physics and Astronomy, Theory Division, Technische Universität Berlin, Hardenbergstrasse 36, 10623 Berlin, Germany}
\date{}

\begin{document}
\maketitle

\begin{abstract}
\noindent\normalsize{
Bacteria often swim in complex environments where surfaces are ubiquitous and rarely flat. Surface topography and curvature can strongly 
affect bacterial motility, with important consequences for surface exploration, adhesion, and biofilm formation. Here, we investigate the 
swimming of a non-tumbling \textit{Escherichia coli} bacterium near an 
undulating no-slip surface using hydrodynamic simulations of a detailed model bacterium. The latter
is described by a rigid spherocylindrical cell body and flexible flagella modeled with the Kirchhoff rod theory, while the surrounding fluid is simulated using the method of
multi-particle collision dynamics. At low curvatures of the sinusoidal surface modulations,
the bacterium exhibits persistent near-surface swimming and clockwise 
trajectories, consistent with the known behavior near flat no-slip walls. As the 
curvature increases, bacteria swimming toward a  ridge can escape from the surface, 
which we use to estimate
a critical curvature where surface detachment is more likely.
At larger curvatures, we find
that the surface geometry promotes oscillatory swimming along the groove direction, which reduces escape opportunities and, therefore, enhances bacterial trapping. 
Indeed, the confinement around the groove reverses the swimming of the bacterium from clockwise to counter-clockwise,
as we demonstrate by two minimal models. Thus our work highlights
the importance of the three-dimensional surface
topography in bacterial surface exploration.}
\end{abstract}

\section{Introduction}
Bacteria live in complex environments where surfaces are ubiquitous.\cite{Donlan2002Biofilms:Surfaces, Hall-Stoodley2004BacterialDiseases,Tuson2013BacteriasurfaceInteractions}
Upon reaching a surface, bacteria can attach, spread, and form complex structures such as biofilms and thereby become more resistant to antiobiotics.\cite{Kimkes2019HowContact,Sharma2023MicrobialTreatment,Liu2024MechanismsBiofilms}
Since bacteria are found in a wide range of biological and biomedical environments, including prosthetic heart valves, catheters, and tissues in the human body, deciphering bacterial motility near surfaces is crucial for controlling biofilm formation.\cite{Sharma2023MicrobialTreatment, Zheng2021ImplicationAdhesion}

Many motile bacteria employ flagella to generate thrust and propel themselves.\cite{2004E.Motion, doi:10.1128/jb.182.10.2793-2801.2000, Lauga2016BacterialHydrodynamics}
The \textit{Escherichia coli} (\textit{E.\ coli}) bacterium, one of the most studied microorganisms, employs relatively rigid helical filaments that are rotated by nanosized motors embedded in the cell membrane.\cite{Berg2003TheFlagella}
When all the filaments rotate in the same direction, a flagellar bundle is formed and the bacterium moves in the bundle direction.\cite{Darnton2007OnColi, Bianchi2023Light-DrivenFormation}
The motion of flagellated bacteria generates a flow field that is well described by Stokes flow singularities, more precisely by a stresslet and a rotlet dipole.\cite{Lauga2009TheMicroorganisms}
At the scale of a single cell, the resulting hydrodynamic interactions with surfaces alter the swimming trajectory.
For instance, it has long been observed that pusher-like swimmers such as \textit{E.\ coli} can align with and become trapped near surfaces, a feature that can be captured, to leading order, by the hydrodynamic interaction between the stresslet and a flat surface.\cite{Spagnolie2012HydrodynamicsApproximations}
This results in both long residence times and higher bacterial densities, which facilitate surface adhesion and therefore biofilm formation.\cite{Drescher2011FluidScattering, Schaar2015DetentionNoise, Junot2022Run-to-TumbleBacteria, Zheng2021ImplicationAdhesion, Bianchi2017HolographicBacteria, Berke2008HydrodynamicSurfaces}
In addition to wall accumulation, \textit{E.\ coli} performs circular trajectories near surfaces, a behavior attributed to the rotlet dipole component of its flow field.\cite{Giacche2010HydrodynamicSurface, Lauga2006SwimmingBoundaries, Mousavi2020WallStudy}
The sense of rotation is determined by the boundary condition and is clockwise (CW) near no-slip surfaces and counter-clockwise (CCW) near free-slip surfaces.\cite{Lemelle2010CounterclockwiseInterface, Hu2015PhysicalSlip}
Such phenomena have been thoroughly investigated both experimentally and theoretically, therefore, flat-surface swimming is now well understood.

In their biological environment, bacteria encounter a wide variety of surface topographies and roughnesses, which can strongly 
affect their motility and exploration capabilities.
In particular, bacterial motility is modified in confined spaces.
For instance, in microchannels or capillary tubes of small radii,
confinement can increase the thrust generated by rotating flagella,
depending on both the channel radius and the cell geometry.\cite{Vizsnyiczai2020AChannels, Acemoglu2014EffectsChannels, Liu2014PropulsionTube}
Furthermore, 
bacteria can wrap their 
flagella
around the cell body to pass through narrow constrictions, while other marine bacteria, upon entering confined spaces, straighten their swimming trajectories to escape confinement.\cite{Yoshioka2026BacteriaWrapping, Lynch2022TransitioningResponse}

Near
curved surfaces, hydrodynamic alignment allows bacteria to follow the surface curvature.
Experimental studies with
pillars have revealed that bacterial density remains higher
near a pillar surface compared to the bulk fluid,
but reduces with decreasing pillar radius.
This highlights
that curvature promotes surface escape.\cite{Sipos2015HydrodynamicWalls, Takaha2023Quasi-two-dimensionalDependence}
Near corrugated surfaces, analytical studies based on
far-field hydrodynamics
have shown that hydrodynamic interactions can decrease wall attraction, depending on the surface wavelength.\cite{Kurzthaler2021MicroswimmersSurfaces}

Although several articles have investigated bacterial motility near various surface geometries, most studies
do not provide a full three-dimensional perspective on the bacterial trajectories.
For instance, in the experimental works of Refs.~\citenum{Perez-Estay2024AccumulationCurvature} and 
\citenum{Mok2019GeometricAccumulation}, the authors studied \textit{E.\ coli} near undulating surfaces, 
which escape from or accumulate at the surfaces depending on the curvature.
However, these studies did not map the full three-dimensional trajectories, in particular, inside grooves.
In contrast, theoretical works have mostly relied on coarse-grained models or far-field hydrodynamics,
which do not account for the detailed
structure of a bacterium or near-field hydrodynamic contributions.\cite{Kurzthaler2021MicroswimmersSurfaces, Das2019ColloidalWalls, Kuron2019HydrodynamicSurfaces}

To further investigate bacterial swimming near curved surfaces, we rely on a numerical model based on the physiology of \textit{E.\ coli}, 
which captures the key features of bacterial motility, including flexible flagella, and which we developed and applied in earlier works.\cite{Vogel2010Force-extensionFlagella,Vogel2013Rotation-inducedFlagella,Adhyapak2015ZippingBody,Adhyapak2016DynamicsRotation,Martin2025E.Study}
The model uses the Kirchhoff rod theory to implement a discrete version of the flexible helical flagella that drive the bacterium
and implements the cell body as a rigid spherocylinder.\cite{Zantop2020}
To resolve the full hydrodynamic flow field generated by the moving \textit{E.\ coli}, we have recently employed the widely used method of multi-particle collision dynamics (MPCD).\cite{Malevanets1999MesoscopicDynamics,Noguchi2007Particle-basedTechniques,Martin2025E.Study}
This method has been extensively used over the past two decades and has proven successful in resolving the hydrodynamics of microswimmers.\cite{Hu2015ModellingColi, Zantop2020, Zottl2018SimulatingDynamics, Zottl2019EnhancedSolutions}

In this article we consider a non-tumbling bacterium swimming close to an undulating surface. The latter is described by
a sinusoidal height modulation, which allows to control the curvature at the top of each hill (ridge) and confinement in the grooves 
by the undulation amplitude. With increasing amplitude, we first observe near-surface swimming on a clockwise trajectory, where the
model bacterium follows the surface modulations. Then, events occur where the bacterium cannot remain near the surface when crossing
a ridge. We provide an estimate for the critical curvature, where the bacterium more likely escapes from the surface. Finally, for large curvatures
we present a mechanism by which the bacterium remains trapped in the groove and effectively moves along it. We clarify the mechanism
by simulating a rotating rod in the groove and a second minimal model, which can be treated analytically.

The article is structured as follows.
In Section\ \ref{sec.modeling} we describe the numerical model bacterium, the parameters used to simulate it, the MPCD method, and
the simulation geometry.
The results are presented in Section\ \ref{sec.results}, where we discuss near-surface swimming on a clockwise trajectory, bacterial
escape, and hydrodynamic trapping. We finish with a summary and conclusions in Section\ \ref{sec.conclusion}.


\section{Modeling and simulation method}
\label{sec.modeling}

We employ the same modeling and simulation framework for \textit{E.\ coli} and its hydrodynamic coupling
to the viscous fluid environment as introduced in Ref.~\citenum{Martin2025E.Study}. We only summarize the main ingredients here and 
refer to Ref.\ \citenum{Martin2025E.Study} for more details.

\subsection{Bacterial model}

\begin{figure}
    \centering
    \includegraphics[width=0.9\columnwidth]{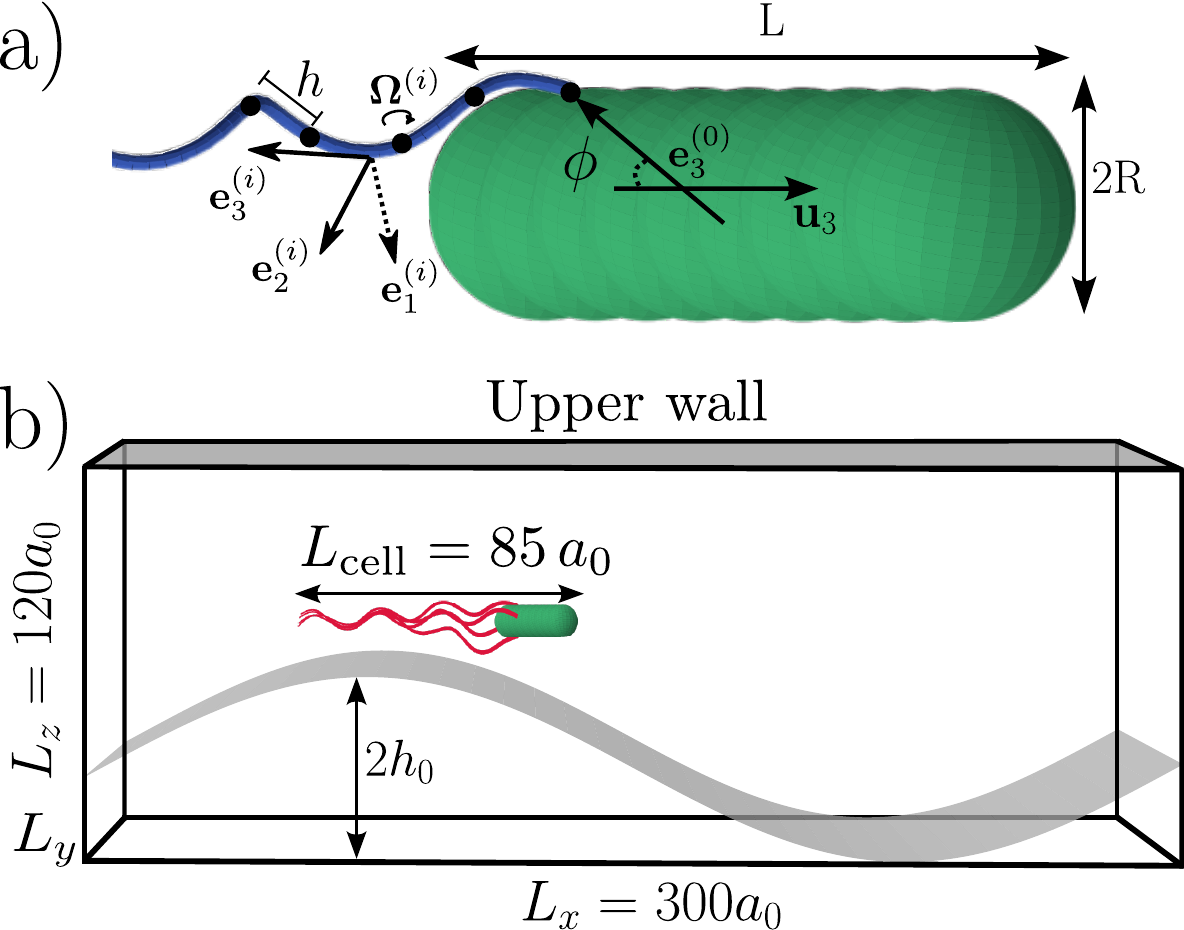}
    \caption{
    (a) Schematic of the modeled bacterium. The cell body has a length $L$ and a width $2R$, with four flagella attached at the rear (only one is shown). The vector $\mathbf{u}_3$ defines the main axis of the cell body. The motor torque $T_m$ is applied along $\mathbf{e}_3^{(0)}$, which encloses
    an angle $\phi$ with $\mathbf{u}_3$, and propagates along the flagellum, transported by the rotational strain vector $\bm{\Omega}^{(i)}$ that acts on the tripods $\{\mathbf{e}_1^{(i)}, \mathbf{e}_2^{(i)}, \mathbf{e}_3^{(i)}\}$.   
    (b) Schematic of the geometry. The space along the $z$-direction is bounded by an upper flat wall and a lower 
    undulating surface located at $z = 120\,a_0$ and $z = 0$, respectively, while periodic boundary conditions are applied along the $x$ and $y$ directions. The wavelength of the undulating
    surface matches the box size $L_x$, and $h_0$ defines its amplitude. The real size of the bacterium within this geometry is also indicated.
   }
    \label{fig:model}
\end{figure}

We treat the bacterial flagellum as a slender body and describe it
by a centerline, which is composed of $N$ beads at positions $\mathbf{r}_i$. We use the Kirchhoff rod theory to model the bending and twisting deformations of the flagellum. The beads are connected by $N-1$ segments of length $h$ that
carry a tripod or \textit{Frenet-Serret} frame, which consists of a
set of orthonormal vectors,
$\{\mathbf{e}^{(i)}_1, \mathbf{e}^{(i)}_2, \mathbf{e}^{(i)}_3\}$.
The vector $\mathbf{e}^{(i)}_3$ represents the local tangent and 
$\mathbf{e}^{(i)}_1$,  $\mathbf{e}^{(i)}_2$
define the cross section of the flagellum as illustrated in Fig.~\ref{fig:model}(a). 
The rotational strain vector $\bm{\Omega}^{(i)}$ determines the configuration of the flagellum via $\partial_s \bm{\Omega}^{(i)} =
\bm{\Omega}^{(i)} \times \mathbf{e}^{(i)}_\nu$ ($\nu=1,2,3$) and deviations $\mathrm{d}\bm{\Omega}^{(i)} = \bm{\Omega}^{(i)} - 
\bm{\Omega}_0$ relative to the ground state 
$\bm{\Omega}_0$ give rise to bending
($\mathrm{d} \Omega^{(i)}_1$, $\mathrm{d}\Omega^{(i)}_2$)
and twisting ($\mathrm{d}\Omega^{(i)}_3$) deformations with local energy
\begin{equation}
f_\text{Kr}(\mathrm{d}\bm{\Omega}^{(i)}) = \frac{A}{2}\left[(\mathrm{d}\Omega^{(i)}_{1})^2 + (\mathrm{d}\Omega^{(i)}_{2})^2\right] + \frac{C}{2}(\mathrm{d}\Omega^{(i)}_{3})^2 \,.
\end{equation}
Here, $A$ and $C$ are the respective bending and twisting rigidities.

In addition, to approximately maintain a constant rod length, we add the local stretching energy:
\begin{equation}
f_\text{st} = \frac{K}{2h} \left(|\mathbf{r}_{i+1} - \mathbf{r}_i| - h\right)^{2} \, .
\end{equation}
with the spring constant $K$.
The total elastic free energy of the flagellum in discretized form then becomes 
$\mathcal{F}(\{\mathbf{r}_i, \phi_i\}) = \sum_i [f_\text{Kr}(\mathrm{d}\bm{\Omega}^{(i)}) + f_\text{st}(\mathbf{r}_i, \mathbf{r}_{i+1}) ]$, where the 
twist angle $\phi_i$ describes a rotation of tripod $i$ about the local tangent.
From this free energy we obtain the force acting on bead $i$ and the twisting torque about segment $i$,
\begin{equation}
\mathbf{F}_\text{el}^{(i)} 
= -\frac{\partial \mathcal{F}}{\partial \mathbf{r}_i}
\qquad \text{and} \qquad
T_\text{el}^{(i)} 
= -\frac{\partial \mathcal{F}}{\partial \phi_i}.
\label{eq.force_torque}
\end{equation}

The cell body is modeled as a line of rigidly connected overlapping spheres, with mass and moment of inertia computed analytically.\cite{Zantop2020} The helical flagella are attached to the cell body, while elastic forces and torques at the anchoring point ensure full mechanical coupling.\cite{Martin2025E.Study}
We represent the rotational flagellar motor that drives the flagellum by a constant torque $T_m$, which acts along a prescribed motor axis $\mathbf{e}^{(0)}_3$, tilted relative to the cell body axis by an angle $\Phi$ [see Fig.~\ref{fig:model}(a)]. The flexibility of the hook is captured by reduced bending and twisting rigidities, allowing the motor torque to be efficiently transmitted at any angle between the motor and the first rod.\cite{Kato2019StructureJoint} Steric interactions 
between filaments, the cell body, and boundaries are modeled using 
the purely repulsive WCA potential, and filament (self-)crossing
is prevented using the method of Ref.~\citenum{Adhyapak2015ZippingBody}.

\subsection{Hydrodynamic coupling to fluid environment}

We model the surrounding fluid using the particle-based method of multi-particle collision dynamics (MPCD) to solve the Navier-Stokes equations in the limit of low Reynolds numbers.
The dynamics of the effective fluid particles on the mesoscale
alternates between streaming and collision steps of total duration $\Delta t$, which are implemented for particle position and velocity by
\cite{Malevanets1999MesoscopicDynamics}
\begin{equation}
\mathbf{r}_i(t+\Delta t) = \mathbf{r}_i(t) + \mathbf{v}_i(t)\Delta t \enspace 
\text{and}
\enspace
\mathbf{v}_i(t+ \Delta t) = \mathbf{u}_\alpha + A(\mathbf{r}_j(t), \mathbf{v}_j(t)).
\end{equation}
During the collision step, the simulation volume is divided into cubic unit cells. Thus, in the second equation, $\mathbf{u}_\alpha$ is the mean velocity of particles in the MPCD cell $\alpha$, and $A(\mathbf{r}_j, \mathbf{v}_j)$ is a collision operator
that conserves total linear and angular momentum in 
cell $\alpha$.

We employ the MPCD-AT algorithm, which uses an Anderson thermostat to maintain a constant temperature.\cite{Gompper2008Multi-ParticleFluids, Noguchi2007Particle-basedTechniques}
However, the standard MPCD-AT scheme does not conserve angular momentum, which can lead to unphysical behavior.\cite{Gotze2007RelevanceSimulations} 
To address this, we use the MPCD-AT$+a$ variant, which restores angular-momentum conservation.
\cite{Noguchi2007Particle-basedTechniques}
The collision operator then reads
\begin{equation} 
A(\mathbf{r}_j, \mathbf{v}_j) =  \mathbf{v}_{i}^{r} - \frac{\sum_{j=1}^{N_{\alpha}}m_{j}\mathbf{v}_{j}^{r}}{\sum_{j=1}^{N_{\alpha}}m_{j}} + \left[\mathbf{I}_{\alpha}^{-1} \sum_{j=1}^{N_{\alpha}}m_{j}\mathbf{r}_{j}^{s}\times(\mathbf{v}_{j}-\mathbf{v}_{j}^{r})\right] \times \mathbf{r}_i^s \, .
\label{eq:collision_step}
\end{equation}
Here, the first term on the right-hand side, $\mathbf{v}_i^{r}$,
is a random velocity drawn from the Maxwell-Boltzmann distribution with variance $\sigma = \sqrt{k_B T_0 / m_0}$,
where $m_0$ is the particle mass and $k_B T_0$ is thermal energy.
The second term ensures conservation of linear momentum.
In the third term, $\mathbf{I}_\alpha$ is the moment-of-inertia tensor of cell $\alpha$, and $\mathbf{r}_i^s$ is the position of particle $i$ relative to the cell center of mass. 
Since the random relative velocities $\mathbf{v}_i^{r}$ change the angular momentum of the cell by $-\Delta \mathbf{L}$, the third term corrects for this change by applying a rotational contribution $\boldsymbol{\omega} \times \mathbf{r}_i^s$ to the particles, where 
$\boldsymbol{\omega} = \mathbf{I}_\alpha^{-1}\Delta \mathbf{L}$
and $\Delta \mathbf{L}$ is the term in square brackets.\cite{Noguchi2007Particle-basedTechniques}

In particular, the flagella are coupled to the fluid via its constituent beads that participate in the collision step, while during MD steps they move with the forces and torques of eqn (\ref{eq.force_torque}).
The no-slip boundary condition at solid surfaces including the cell body is enforced using a bounce-back rule.\cite{Zottl2018SimulatingDynamics}
Virtual particles are introduced in partially filled cells that overlap with the bounding surfaces to 
keep the number of particle per cell constant on average.
By participating in the collision step, they enhance the enforcement of no-slip boundary conditions.\cite{Gompper2008Multi-ParticleFluids}
Finally, when fluid particles collide with the cell body through the bounce-back rule, they transfer momentum and the linear and angular momentum of the cell body is updated.
We
employ the velocity Verlet scheme to integrate Newton's equations for both the cell body and the flagella.

\subsection{Simulation geometry}

In the following, we consider two bounding surfaces along the vertical $z$ direction [see Fig.\ \ref{fig:model}(b)]:
an upper flat wall parallel to the $xy$ plane, and a lower sinusoidal surface defined by the height modulations
\begin{equation}
z(x,y) = h_0  \left[1 + \sin\left(\frac{2\pi x}{\lambda}\right) \right] \, ,
\end{equation}
where $h_0$ and $\lambda$ are the amplitude and wavelength, respectively. Periodic boundary conditions are applied in the $x$ 
and $y$ directions.

For later use, we mention that the maximum local curvature of the undulating surface in the valleys and the hills amounts to 
\begin{equation}
\label{max_curvature}
    \kappa_\text{max} = 4 \pi^2 \frac{h_0}{\lambda^2}
\end{equation}
This expression follows from the standard formula for the curvature of a graph $y=f(x)$, 
$\kappa= f(x)''/(1+f(x)')^{3/2}$, which we evaluated at the minima and maxima of $z(x,y)$. Note that the curvature is linear in the amplitude 
$h_0$ and quadratic in $\lambda^{-1}$. 

\subsection{Simulation parameters}

MPCD provides a natural set of units to describe the system. 
Typically, one takes the lattice constant $a_0$,
the particle mass $m_0$, and the thermal energy $k_B T_0$ as respective units for length, mass, and energy.
Furthermore, one derives units for time $\tau_0 = a_0 \sqrt{m_0 / k_B T_0}$, viscosity $\eta_0 = \sqrt{m_0 k_B T_0}/a_0^2$, and density $\rho_0 = m_0 / a_0^3$. We use them to define the parameters in our simulations.
The particle density is set to $\rho = 10\, m_0/a_0^3$ and the time step is $\Delta t = 0.05\,\tau_0$, which yields a viscosity $\eta = 7.5\,\eta_0$.\cite{Noguchi2008TransportTechniques} 
Finally, at room temperature $k_B T_0 = 4.13\,\mathrm{pN\,nm}$.

To incorporate the \textit{E.\ coli} model into the MPCD simulations, we choose $a_0 = h = 0.1\,\mu\mathrm{m}$. 
Thus, the discretization of the flagellar filament
is chosen such that, on average, each flagellar bead occupies one
MPCD cell, which ensures a proper resolution of the local
anisotropic friction.\cite{Martin2025E.Study}
Each flagellum consists of $N = 80$ beads, so the corresponding contour length is $8\,\mu\mathrm{m}$. 
The bead mass is set to $10\,m_0$, which ensures smooth velocity changes during a collision step and matches the solvent density.\cite{Gompper2008Multi-ParticleFluids, Ripoll2005DynamicDynamics}
The cell body is constructed from nine overlapping spheres. To
obtain a smooth cell surface,the
spheres are separated by $d = 1.9\,a_0$ and have radius $R = 4.4\,a_0$, which gives a cell-body length of $L = 24\,a_0$
in agreement with experimental values.\cite{doi:10.1128/jb.182.10.2793-2801.2000}
We anchor four flagella symmetrically on the cell-body surface with a distance of $6.4\, a_0$ from one end. Thus, considering
the helical shape of the flagella, the total cell length consisting of flagella and cell body amounts to
$L_\text{cell} 
\approx
85\,a_0$.

To reproduce a realistic flagellar thickness via the WCA potential, we use for its length parameter
$\sigma = 0.4\,a_0$, which sets the diameter of the flagellum.
For the flagellum-surface interactions, we use $\sigma = 0.5\,a_0$ and always $\epsilon = k_B T_0$.
The rotational strain vector of the helical ground state
is set to 
$\bm{\Omega}_0 = \{0.0, 0.13, -0.211\}\,a_0^{-1}$, which yields the experimentally observed flagellar helix.
This parameter can be tuned to model different configurations of the flagellar polymorphism, as shown in Refs.~\citenum{Adhyapak2016DynamicsRotation} and \citenum{Martin2025E.Study}.
The bending and twisting rigidities are set to $A = 13310\,k_B T_0 a_0$ and $C = 8450\,k_B T_0 a_0$, respectively, in agreement
with experimental measurements.\cite{Darnton2007OnColi, Darnton2007Force-extensionTransformations} 
The hook rigidities are chosen as
$A_\text{hook} = 2.5\,k_B T_0 a_0$ and $C_\text{hook} = 4840\,k_B T_0 a_0$. To prevent rod elongation, a large spring constant
$K = 10^5\,k_B T_0 / a_0$ is used.

The motor torque is set to $T_m = 820\,k_B T_0$ consistent with the experimental values for the stall torque and previous simulations, 
which gives a swimming velocity of $v \approx 0.03 v_0$.\cite{Berg2003TheFlagella, Adhyapak2016DynamicsRotation, Martin2025E.Study}. 
We set the tilt angle to
$\Phi = 55^\circ$, which results in a counter rotation of the cell body five times slower than the flagellum.\cite{Darnton2007OnColi} 
Finally, we define the time scale $\tau_c = 2650 \tau_0$, which is
the average time it takes for the model $E.\ coli$ to run through its total length.

The simulations are performed in a box of dimensions $L_x = 300\,a_0$, $L_y = 200\,a_0$ and $L_z = 120\,a_0$, with the upper 
wall located at $z = 120\,a_0$. The wavelength of the sinusoidal surface is set to $\lambda = L_x$, so that $h_0$ is the only parameter 
varied to control the surface curvature. An illustration of the system is shown in Fig.~\ref{fig:model}(b).

\section{Results}
\label{sec.results}

In the successive subsections, for increasing surface-undulation amplitude we present our results on near-surface swimming on 
a clockwise trajectory, on bacterial escape, and hydrodynamic trapping in a groove.

\subsection{Low curvature regime: Near surface swimming and clockwise trajectory}
\label{sec:Low_curvature}
\begin{figure}
    \centering
    \includegraphics[width=0.95\columnwidth]{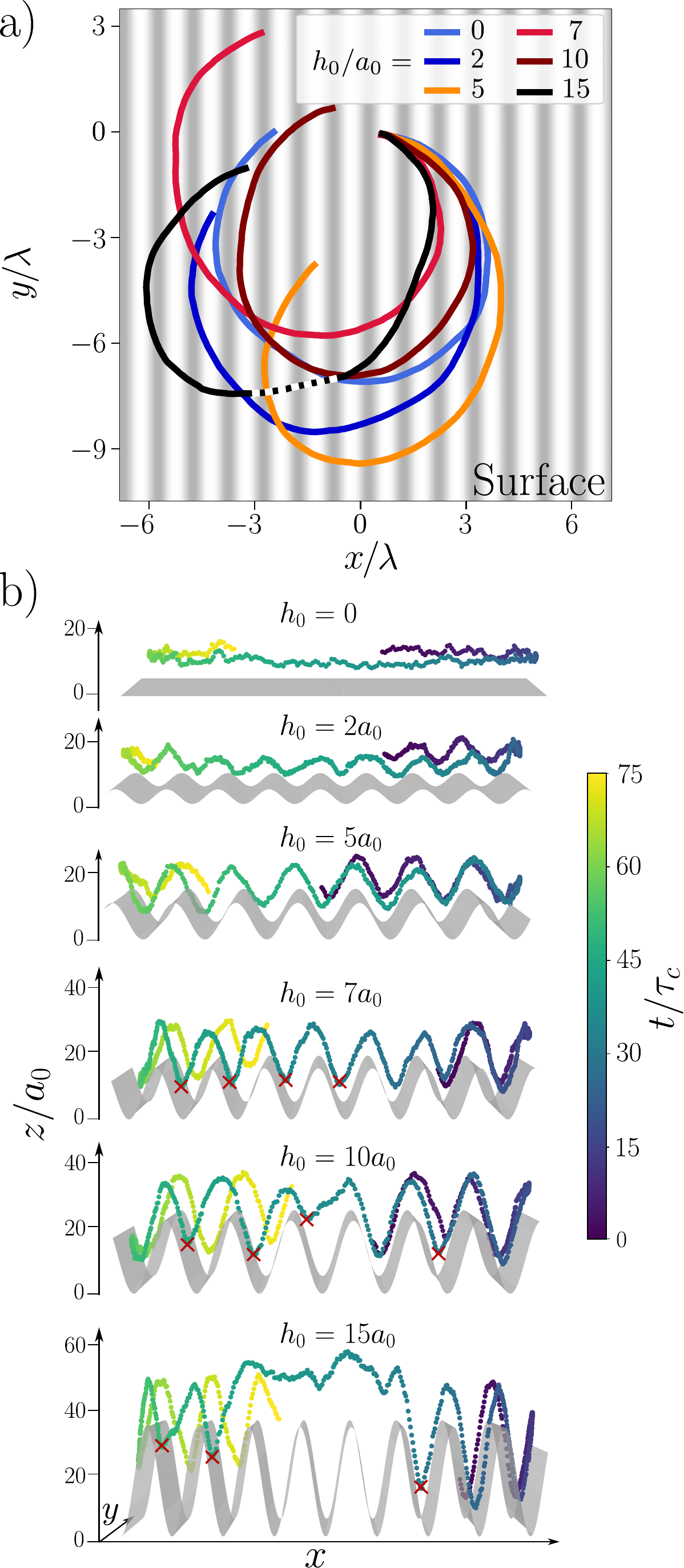}
    \caption{Representative trajectories for several undulation amplitudes
    $h_0$, shown from different perspectives.
    (a) 2D top view, with $x$ and $y$ given in units of undulation wavelength $\lambda$. The gray shading indicates the surface undulation, 
      with the white stripes corresponding to grooves.
    The dashed segment on the black curve highlights the bacterium swimming above the surface, as discussed in the main text.
    (b) 3D side view. Color encodes time, and red crosses highlight collisions of the cell body with the hills of the 
    undulating surface. The trajectories start in the background, circle to the front, and move to the background again.
    Note that $x$ and $z$ are scaled differently, which overemphasizes the undulations.}
    \label{Trajectories}
\end{figure}

In the beginning of a simulation, we orient the bacterium along the $x$-direction and close to the lower surface, with 
the flagellar bundle already formed.
In the presence of surface undulations, the bacterium starts at $x = \lambda/2$, half way between groove and
hill and always tilted about the $y$ axis towards the groove.
We simulated long trajectories with duration of $75\,\tau_c$, so the bacterium moves 75 times its total length.
We varied the undulation amplitude in steps from $h_0 = 0$ (the flat case) to $h_0 = 15\,a_0$.
Figure\ ~\ref{Trajectories} reports representative trajectories from different perspectives.

Due to the hydrodynamic stresslet flow field, generated by an extensile force dipole, the bacterium is hydrodynamically attracted to a surface. Therefore, \textit{E.~coli} accumulate near a surface with a density higher than in the bulk fluid. \cite{Berke2008HydrodynamicSurfaces, Junot2022Run-to-TumbleBacteria}
Moreover, the rotating bundle and counter rotating cell body form a so-called rotlet dipole, which, in the vicinity of a surface, leads to circling in CW direction, as observed in Fig.~\ref{Trajectories}(a). 
Deviations from a perfect circle are due to rotational noise and the surface undulations.
For $h_0 \neq 0$, the bacterium is tilted and swims in the direction of the groove. 
Upon crossing a groove, the bacterium experiences a hydrodynamic torque due to its extensile stresslet interacting with the curved 
surface, which reorients the bacterium \cite{Berke2008HydrodynamicSurfaces}.
Furthermore, for increasing undulations, steric reorientation by the curved surface becomes more and more relevant.
When crossing the ridge, only the hydrodynamic torque acts. Thus, only for small $h_0$,  this torque is sufficient to guide the bacterium along 
the surface, while for larger $h_0$ the bacterium leaves the surface and eventually escapes, which we investigate more thoroughly in the
next subsection. In the following, we describe the occuring trajectories in more detail.

At $h_0 = 2\,a_0$ and $5 \,a_0$,
the bacterium consistently follows a curved path, which generates a trajectory with sinusoidal height variation
aligned with the undulating surface [see Fig.\ \ref{Trajectories}(b)]. This is
reminiscent of the experimental observations in Ref.~\citenum{Perez-Estay2024AccumulationCurvature}. The circular trajectories behave 
similar to those observed near a flat surface. Increasing the amplitude to $h_0 = 7\,a_0$ alters the ability of the modeled bacterium to 
swim persistently close to the undulating surface. Upon crossing a ridge, the bacterium visibly detaches from the surface and, being reoriented 
by the hydrodynamic torque, then collides with the next hill.
There, it reorients along and reattaches to the surface. We have indicated a few examples of such events
by red crosses in Fig.~\ref{Trajectories}(b).
This repeated detachment-attachment behavior clarifies
that hydrodynamic interactions alone are no longer sufficient to ensure the bacterium adheres perfectly to the surface.
For $h_0 = 10\,a_0$ and $h_0 = 15\,a_0$, the detachments 
become more pronounced
(see Video 1 in the supplementary material).
In both cases, at time $t \approx 35,\tau_c$, the bacterium cannot follow the undulating
surface and swims above the undulations not diving into the grooves.
For $h_0 = 15\,a_0$, this leads to sustained motion above the undulating
surface starting at $t \approx 15\,\tau_c$, which explains the relatively straight trajectory observed in the top view of Fig.~\ref{Trajectories}(a).

\begin{figure}
    \centering
    \includegraphics[width=1\columnwidth]{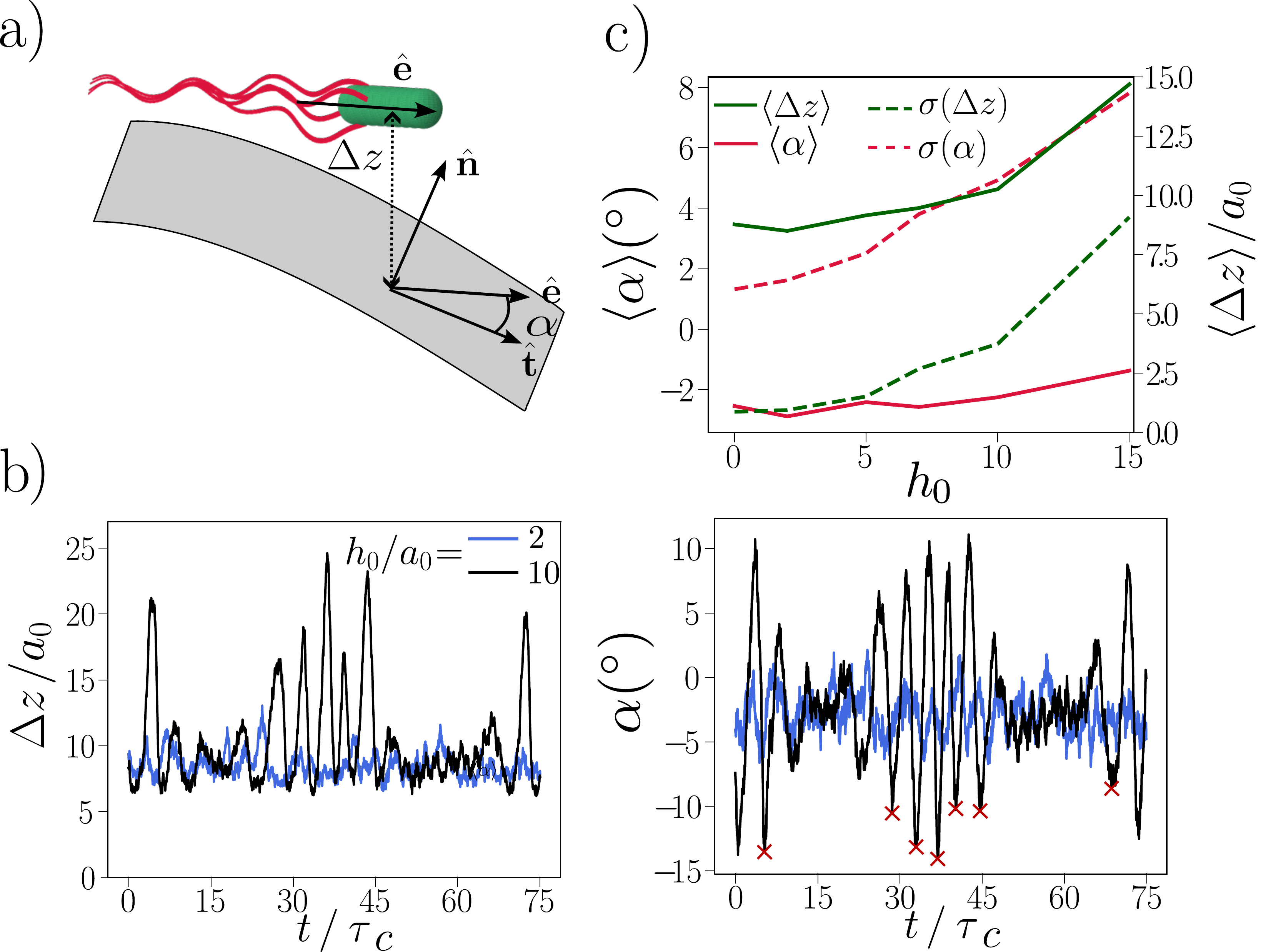}
    \caption{
    (a) Schematic illustrating the definition of the analyzed quantities: the vertical distance $\Delta z$
    between the center of mass of the cell body and the undulating surface, and the angle $\alpha$ (measure in $^\circ$) between the bacterial orientation vector $\hat{\mathbf{e}}$ and the local tangent vector $\hat{\mathbf{t}}$ at the surface, which lies in the plane defined by $\hat{\mathbf{e}}$ and the surface normal $\hat{\mathbf{n}}$.
    (b) Time series of $\Delta z$ and $\alpha$ for trajectories with $h_0 = 2\,a_0$ and $h_0 = 10\,a_0$. Red crosses highlight collisions of the cell body with the hills of the undulating surface.
    (c) Time averages $\langle \alpha \rangle$, $\langle \Delta z \rangle$ (solid lines) and standard deviations $\sigma(\alpha)$, $\sigma(\Delta z)$ plotted \emph{vs.} $h_0$.}
    \label{alpha_delta_z}
\end{figure}

To quantify these observations, we determine the orientation and distance of the bacterium
relative to the surface. 
Alignment with the surface is quantified by an angle $\alpha$, which we define
as the angle between 
the orientation vector $\hat{\mathbf{e}}$ of the bacterium
and a local tangent vector $\hat{\mathbf{t}}$ of the surface at the cell-body position $x_{\mathrm{cell}}$,
as illustrated in Fig.~\ref{alpha_delta_z}(a). 
The local tangent $\hat{\mathbf{t}}$ is tangential to a curve which is the intersection of the plane defined by the surface normal 
$\hat{\mathbf{n}}$ and the orientation vector $\hat{\mathbf{e}}$.
The orientation vector
of the bacterium is defined as the unit vector pointing from the flagellar bundle to the center of mass of the cell body.
This definition minimizes fluctuations due to the wobbling dynamics of the cell body. 
Furthermore, we consider the vertical distance of the bacterium from the undulating surface,
$\Delta z = z_{\mathrm{cell}} - z(x_{\mathrm{cell}}, y)$.

Figure~\ref{alpha_delta_z}(b) shows the time series of $\alpha$ and $\Delta z$ for two cases: $h_0 = 2\,a_0$ and $h_0 = 10\,a_0$. 
For the small amplitude $h_0 = 2\, a_0$, both the cell-surface distance and the relative orientation remain relatively constant.
When swimming along the undulating surface, the bacterium reorients mainly by the hydrodynamic torque. Since the reorientation is not instantaneous, this causes weak oscillations in $\Delta z$ and $\alpha$, as a closer look shows, with superimposed small fluctuations 
caused by thermal noise.

For $h_0 = 10 \, a_0$, much larger fluctuations occur in both quantities.
The maximum distance from the surface at $t \sim 33\,\tau_c$ corresponds to the bacterium passing the groove without diving into it [see Fig.\ \ref{Trajectories}(b)].
The peaks in $\alpha$ highlight successive detachments from the surface at the ridges ($\alpha > 0$) and collisions with the subsequent hill ($\alpha < 0$). 
Note that in the time intervals from $t \approx10\,\tau_c$ to $20\,\tau_c$ and $t \approx 55\,\tau_c$ to $65\,\tau_c$, the bacterium is mainly oriented along the $y$ direction, consequently, the surface ahead appears flat, and fluctuations are much reduced.

To obtain a complete view, we report in Fig.~\ref{alpha_delta_z}(c) the time averages $\langle \alpha \rangle$, $\langle \Delta z \rangle$ (solid lines) as well as the standard deviations $\sigma(\alpha)$, 
$\sigma(\Delta z)$ as functions of $h_0$.
For a flat surface ($h_0 = 0$), the bacterium always tilts slightly
towards the surface, as indicated by $\alpha < 0$.\cite{Bianchi2017HolographicBacteria, Bianchi20193DInterface} 
Furthermore, the small $\sigma(\Delta z)$ is in agreement with swimming at nearly constant height, as expected for a flagellated bacterium near a flat wall.\cite{Giacche2010HydrodynamicSurface} 
Interestingly, as $h_0$ increases $\langle \alpha \rangle$ remains nearly constant indicating that the bacterium tilts away and towards
the undulating surface, while $\langle \Delta z \rangle$ increases slightly until $h_0 \approx 10 \, a_0$ and then more strongly. This agrees
with attachment-detachment events becoming more pronouced at $h_0 \approx 10 \, a_0$. The steadily increasing standard
deviations for both quantities indicate the increasing variations around the mean value. In conclusion, the attachment-detachment events
indicate that hydrodynamic reorientation at the ridges is no longer sufficient to prevent the bacterium from leaving the surface.


\subsection{Analysis of the escape from undulating surfaces}
\label{subsec.surface_escape}

\begin{figure}
    \centering
    \includegraphics[width=0.95\columnwidth]{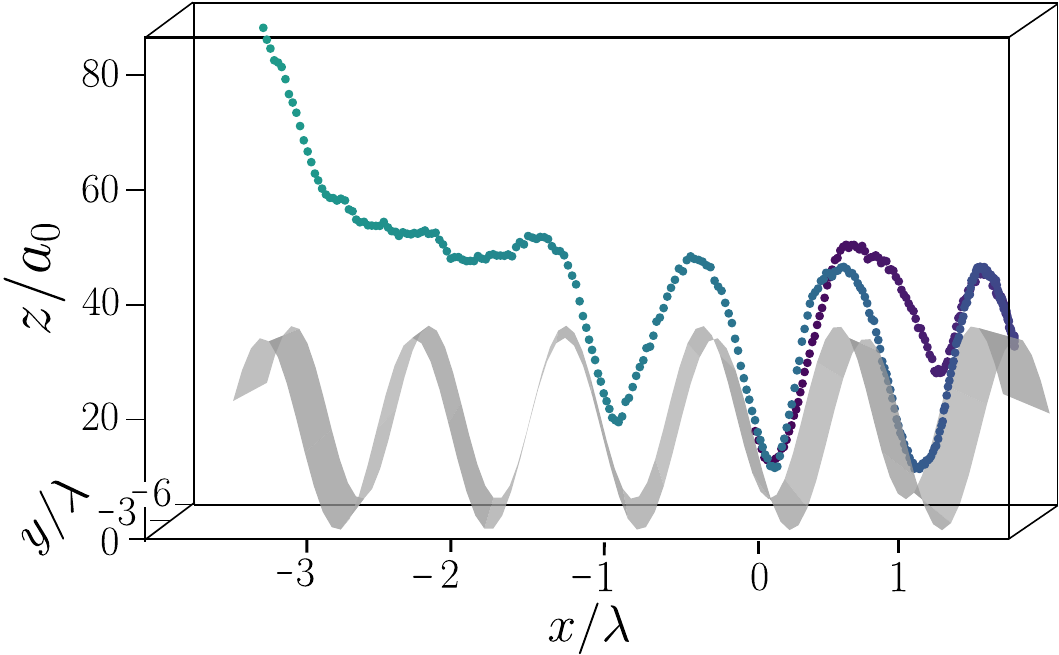}
    \caption{Representative trajectory of a bacterium escaping an undulating surface with amplitude $h_0 = 15a_0$
    after crossing a ridge. Compared to Fig.~\ref{Trajectories} the trajectory is shorter
    but the color coding of time is the same.
}
    \label{fig:escape}
\end{figure}

\begin{figure}
    \centering
    \includegraphics[width=0.90\columnwidth]{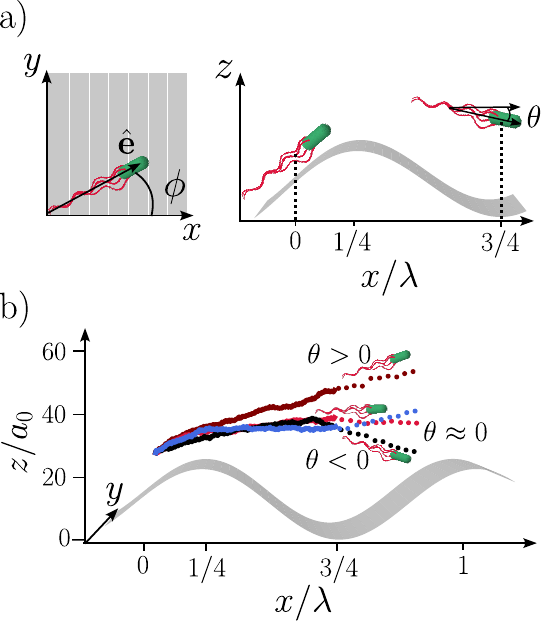}
    \caption{(a) Schematic illustrating the setup employed to quantitatively analyze escape events. 
     Left: Definition of $\phi$, the angle between the direction of the undulating surface along the $x$ axis and the bacterial orientation 
     $\hat{\mathbf{e}}$. 
     Right: The bacterium starts close to the surface at the steepest slope of the undulation at $x=0$ with initial orientation parallel to the 
     surface ($\alpha=0$) and pointing along the undulations ($\phi=0$). At the position $x= 3\lambda/4$ the angle $\theta$ between 
     $\hat{\mathbf{e}}$ and the horizontal plane is determined.
     (b) Four bacterial trajectories for $h_0 = 10\,a_0$ starting with the initial conditions introduced in (a). The solid lines show the bacterial 
     trajectories up to the position $ x= 3\lambda/4$  and the dotted lines indicate the continued trajectories.}

    \label{fig:exemple_traj_theta}
\end{figure}

Bacteria are hydrodynamically attached to flat surfaces and can only escape by tumbling or rotational diffusion of the long axis,
Non-tumbling bacterical strains are observed to remain trapped near surfaces for extended periods, indicating that thermal noise alone is generally insufficient to induce an escape.
In the previous subsection, we observed an attachment-detachment pattern for surfaces with an undulation amplitude $h_0 = 15\,a_0$. Now, for the same amplitude, 
Fig.~\ref{fig:escape} also shows a complete escape of the bacterium from the undulating surface
(see Video 2 in the supplementary material).
This suggests that thermal fluctuations in the bacterial orientation play a subtle role here for the escape, unlike flat surfaces.
To analyze the bacterial escape further, we note that the bacterium moving along the undulating surface experiences
a spatially varying curvature, which depends both on $h_0$ and on the orientation of the bacterium with respect to the 
surface undulations along the $x$ axis. When the bacterium moves with an angle $\phi$ relative to the $x$ axis
[see Fig.\ \ref{fig:exemple_traj_theta}(b)], it experiences undulations with an effective wavelength 
$\lambda_\phi = \lambda / \cos \phi $ and the maximum curvature mentioned in  Eq.\ (\ref{max_curvature}) becomes
\begin{equation}
\kappa = 4\pi^2\cos^2\phi \, h_0/\lambda^2 \, .
\label{eq.curvature}
\end{equation}
Therefore, the angle $\phi$ at which the bacterium crosses a ridge is a key parameter controlling bacterial escape,
and it quadratically modifies the maximum curvature.

To estimate the critical curvature $\kappa_0$, beyond which an escape occurs predominantly,
we performed simulations, in which the center of mass of the model \textit{E.~coli} is positioned at $x = 0$, 
the location of steepest descent. Furthermore, the bacterium is always oriented tangentially to the surface ($\alpha =0$)
and with an angle $\phi$ relative to the $x$ axis [see Fig.\ \ref{fig:exemple_traj_theta}(a)].
In Fig.~\ref{fig:exemple_traj_theta}(b) we show four representative trajectories for $h_0 = 10\,a_0$, with initial orientation along the 
undulation direction ($\phi = 0$) and then monitor the bacterial orientation (angle $\theta$) with respect to the horizontal plane at 
$x/\lambda = 3/4$, where a groove is located [see Fig.\ \ref{fig:exemple_traj_theta}(a)].
As mentioned above, orientational noise significantly affects the possible trajectories: 
in some cases the bacterium orients away from the surface and seems to escape (brown curve), 
while in others it reorients and swims towards the surface (black curve). We already showed these two types 
of trajectories for $h_0 = 15 \, a_0$ in Figs.\ \ref{Trajectories}(b) and \ref{fig:escape}. However, we also observe the ambiguous behavior 
(red curve), where the bacterium is oriented along the horizonal ($\theta \approx 0$) and, therefore, can either stay at or escape from the 
surface (red curve).
Note that the classification of escape, formulated here, is just a snapshot in time. Escaped bacteria can still return to the surface 
and \emph{vice versa}. The blue trajectory in Fig.~\ref{fig:exemple_traj_theta}(b), while classified as ambiguous ($\theta \approx 0$) at 
$x/\lambda = 3/4$, will then escape as the dotted line shows. A careful inspection of the black trajectory just reveals a possible escape before it 
turns towards the surface at $x/\lambda = 3/4$. However, monitoring the orientation angle $\theta$ above the groove will help us to give some rough measure, when bacteria escape, on average.

Accordingly, to determine a value for the critical curvature $\kappa_0$, we average the orientation angle $\theta$ over 50 independent simulations.
As we already stated, to monitor the orientation, the location of the groove is chosen.
In particular, when approaching the groove by diving into the valley, the bacterium only experiences the hydrodynamic torque, while beyond the groove the bacterium might be reoriented during collisions with the surface. 
This then produces positive values of $\theta$ that could be misinterpreted as an escape.
Because the maximum curvature, a bacterium experiences along its path,
depends on both the amplitude $h_0$ and the in-plane orientation $\phi$, we vary both parameters.
Note that a larger $\phi$ also means a larger path length from the starting point to the neighboring groove. We estimated that
the circling of the bacterium close to surfaces does not alter the in-plane orientation noticeably, so $\phi$ is a controllable parameter.

\begin{figure}
    \centering
    \includegraphics[width=0.95 \columnwidth]{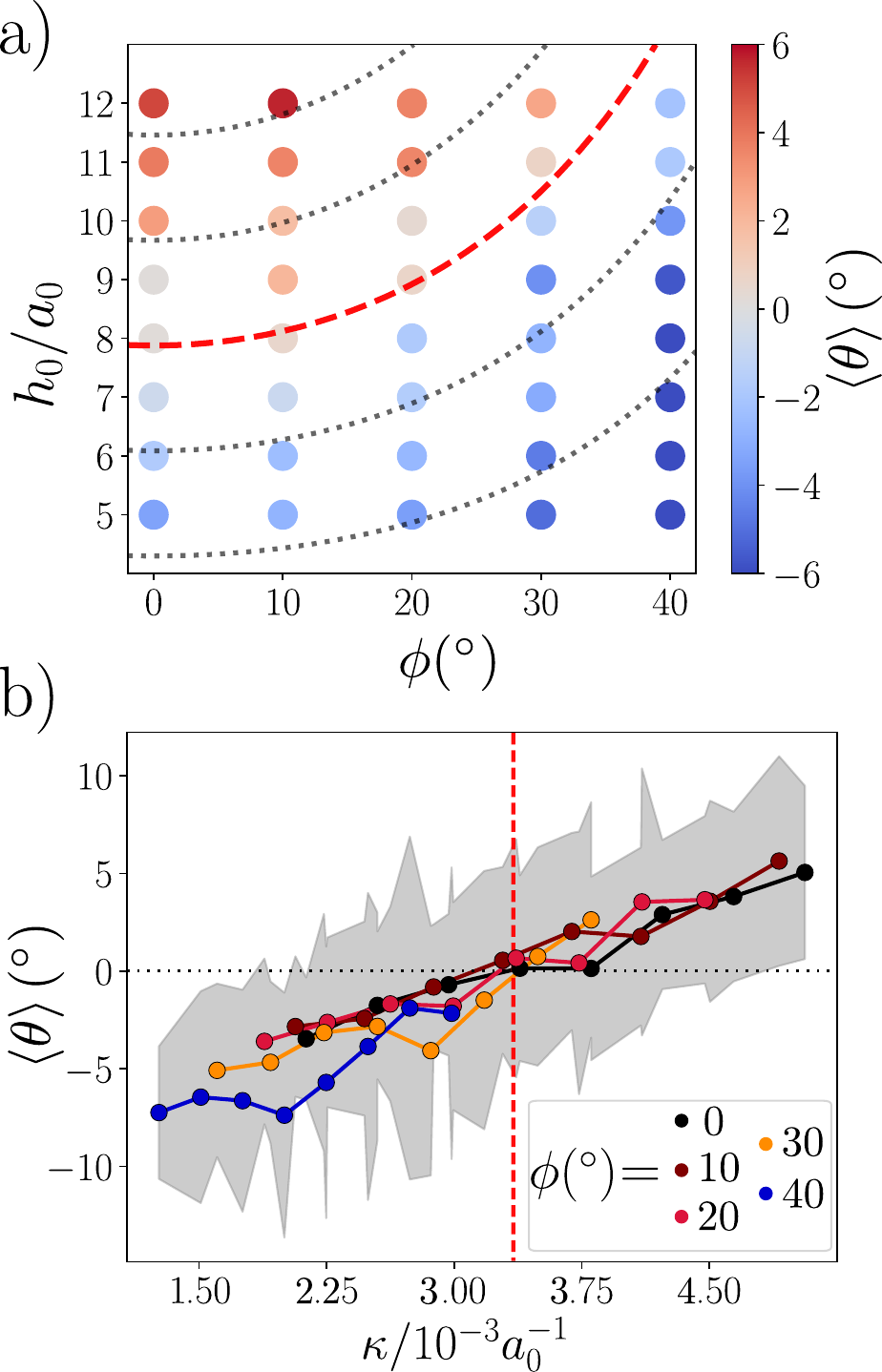}
    \caption{
    (a) State diagram of the mean escape angle $\langle \theta \rangle$ as a function of the undulation
    amplitude $h_0$ and the initial in-plane angle $\phi$. 
    The red dashed line of the critical curvature $\kappa_0=3.4\times 10^{-3}\, a_0^{-1}$ was fitted using
    Eq.\ (\ref{eq.curvature}) such that it hits the points with
    $\langle \theta \rangle \approx 0$. The black dashed lines belong to constant curvature values $\kappa \ne \kappa_0$.
    (b) Mean escape angle $\langle \theta \rangle$ plotted \emph{vs} $\kappa$ for different in-plane angles $\phi$. The same data 
    as in (a) is used,
and the standard deviation is shown as a gray band. The 
red vertical dashed line indicates
$\kappa_0$.
}
    \label{fig:phasediagram}
\end{figure}

In Fig.~\ref{fig:phasediagram}(a), we present a state diagram in the parameters  $h_0$ \emph{versus} $\phi$, where the mean orientation angle is shown by the color code.
The blue and red dots correspond, respectively, to bacteria that are more likely to reorient towards the surface
($\langle \theta \rangle <0$) or to escape ($\langle \theta \rangle >0$) and the gray points correspond to $\theta \approx 0$.
To connect to the maximum curvature, lines of constant curvature are indicated by dashed lines. In particular, 
the red dashed line is chosen such that it roughly hits the dots with zero $\langle \theta \rangle$, which gives a critical curvature of 
$\kappa_0=3.4\times 10^{-3}\, a_0^{-1}$.  Along the other dashed lines, the data points also
have approximately the same value of $\langle \theta \rangle$.
This all is consistent with the idea that escape is governed by the effective curvature of the surface, which the bacterium experiences along its path.

Figure\ \ref{fig:phasediagram}(b) reveals an approximately linear relationship between $\langle \theta \rangle$ and
$\kappa$, independent of $\phi$, which again highlights the role of surface curvature for bacterial escape. 
The critical curvature $\kappa_0$, indicated by the red dashed line, separates again a positive mean orientation angle with more likely 
escape ($\langle \theta \rangle >0$) from bacteria staying at the surface ($\langle \theta \rangle < 0$).
We compared our findings with the work in Ref.~\citenum{Sipos2015HydrodynamicWalls}, where experiments with
$E.\ coli$ swimming near pillars of different radii were performed.
The authors observed that for pillars of $50 \, \mu\mathrm{m}$ nearly $50\, \%$ of the bacteria escaped, corresponding to 
a curvature $\kappa = 2 \times 10^{-2}\,\mu\mathrm{m}^{-1}$. Interestingly, this value is 
close to
our critical curvature which amounts to 
$\kappa_0 = 3.4 \times 10^{-2}\,\mu\mathrm{m}^{-1}$ for $a_0 = 0.1 \mu\mathrm{m}$.
We also note that, despite performing 50 simulations for each value of $\langle \theta \rangle$,
the standard deviation indicated by the gray area in Fig.\ \ref{fig:phasediagram}(b) remains large, emphasizing that 
orientational noise is a determining parameter.

Finally, we compare the individual trajectories in Fig.\ \ref{Trajectories}(b) with the state diagram in Fig.~\ref{fig:phasediagram}(a). 
For example, for $h_0 = 10 \, a_0$ the state diagram predicts a very likely escape for $\phi \leq 20^{\circ}$, while the trajectory
in Fig.\ \ref{Trajectories}(b) does not show any escape. The main reason is that the idealized starting conditions introduced in 
Fig.\ \ref{fig:exemple_traj_theta}
to determine the angle $\theta$ are not generally realized in a simulated trajectory, meaning the angle $\alpha$ with respect to the surface
and the height above the surface are strongly varying. Nevertheless, the state diagram clearly illustrates
the important role the maximum curvature plays for a bacterium moving close to an undulating surface and potentially escaping from it.

\subsection{High curvature regime: Hydrodynamic trapping in the grooves}

\subsubsection{Phenomenology}

\begin{figure}
    \centering
    \includegraphics[width=1\columnwidth]{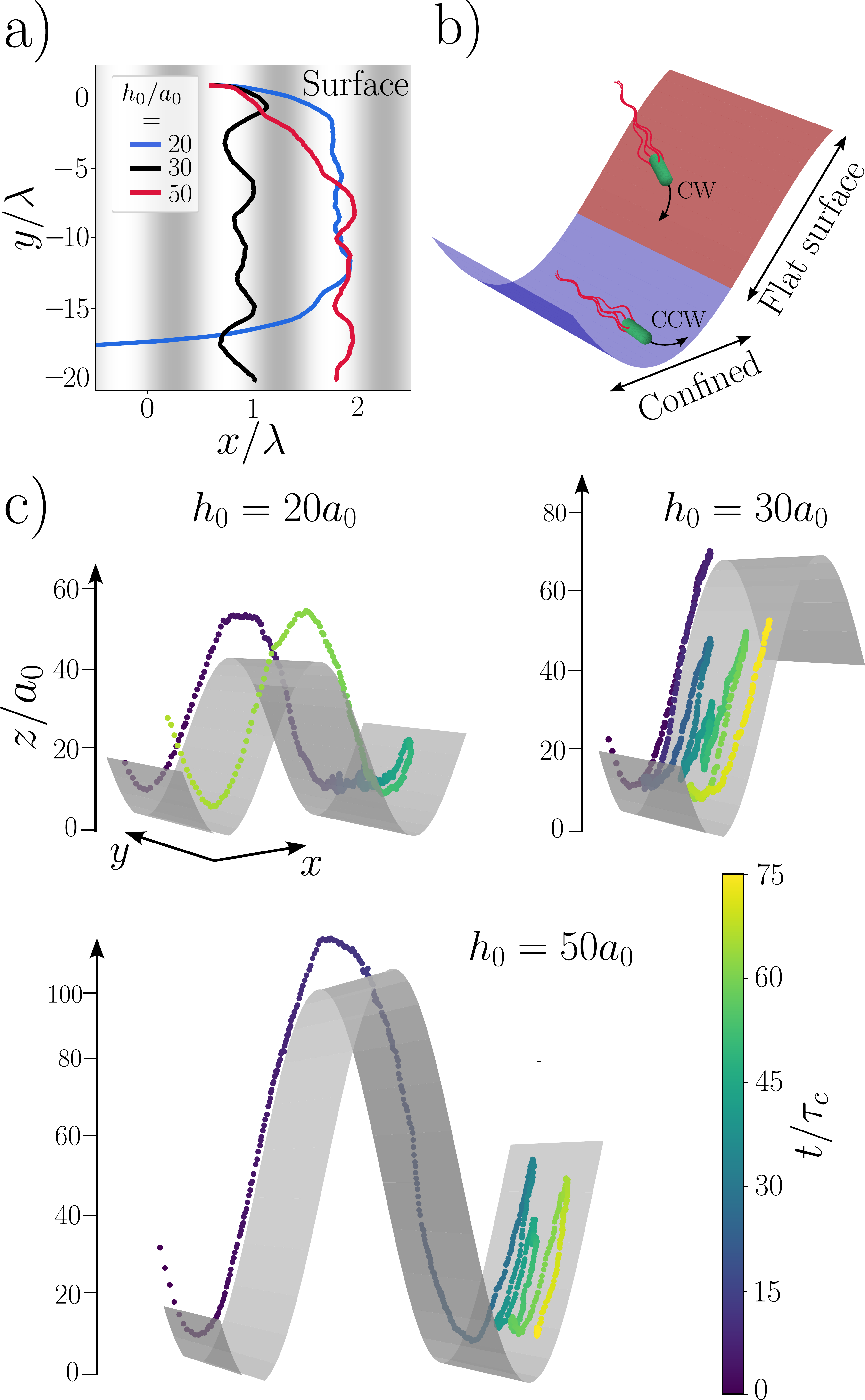}
    \caption{
    Representative trajectories as in Fig.~\ref{Trajectories} but for larger undulation amplitudes $h_0$, shown from different perspectives.
    (a) 2D top view of three trajectories starting at the origin. The grooves correspond to the white stripes.
    (b) Schematic illustrating
    the CCW and CW motions of the bacterium in different parts of the undulating surface, which is
     the origin of the oscillatory motion within the groove. 
    (c) 3D side view of the three trajectories from (a).
    Color indicates time evolution. 
    For $h_0 = 20 \, a_0$, the trajectory is truncated 
    once the initial groove is reached again
    to emphasize the relevant feature.}
    \label{fig:ex_traj_oscillation}
\end{figure}
In continuation of the previous section, we performed simulations with increasing undulation amplitude $h_0 = 20, 30$, and $50\,a_0$.
The bacterium is initialized with the same position and orientation as described at the beginning of Sec.\ \ref{sec:Low_curvature}.
From the state diagram in Fig.~\ref{fig:phasediagram} and for increasing $h_0$, we would expect successful  bacterial escapes 
even at larger incident angles $\phi$ upon reaching a ridge.
However, interestingly, we find long routes where the bacterium swims predominantly along the $y$-direction within a groove. 
So, the typical clockwise (CW) trajectory near a no-slip surfaces is no longer observable. This is clearly shown in Fig.\ \ref{fig:ex_traj_oscillation}(a)
as a 2D top view for trajectories, starting at the position of steepest descent at $x =\lambda/2$, for several values of $h_0$. The trajectories either 
cross the ridge ($h_0 = 20$, $50a_0$) or do not reach it
($h_0=30a_0$) but ultimately end up moving, for some while, parallel to the groove in an oscillatory fashion.
We will argue that the physical mechanism underlying this observation is purely hydrodynamic.
When the bacterium swims along the  ``planar part'' of the undulating surface around the area of steepest descent [see Fig.\ \ref{fig:ex_traj_oscillation}(b)], it experiences hydrodynamic interactions with the
no-slip surface and thus follows a CW trajectory.\cite{Lauga2006SwimmingBoundaries}
However, upon entering a groove [see Fig.~\ref{fig:ex_traj_oscillation}(b)], the confinement in the groove reverses the 
sense of rotation and the bacterium swims on a counter-clockwise (CCW) trajectory.
In Fig.~\ref{fig:ex_traj_oscillation}(c), we present 3D representations of the trajectories for the three values of $h_0$,
which we describe in more detail in the following.

At $h_0 = 20 \, a_0$, the bacterium first traverses the groove and reaches the ridge with a large angle $\phi \approx 60^\circ$. In accordance with the state diagram in Fig.\ \ref{fig:phasediagram}(a), at such a large $\phi$ the bacterium
experiences a sufficiently small curvature so that it crosses the ridge without escaping from the surface. 
It continuous swimming CW along the undulating surface
and reaches the second groove at $t \approx 10 \tau_c$. 
From this time on, it remains approximately aligned with the $y$-axis for about 
a period of $40 \tau_c$.
This behavior clearly contrasts with our observations in Section~\ref{subsec.surface_escape},
where the bacterium consistently exhibits CW motion for $h_0 \leq 15 \, a_0$. At time $t \approx 50 \tau_c$,
the bacterium eventually crosses the groove, and the CW trajectory resumes. We note that, at 
a height equal to the cell-body radius,
the groove has a width of
ca.
 $50\, a_0$. This width is much larger than the cell-body radius, suggesting that steric interactions with the undulating surface are not sufficient to impose alignment along the $y$-direction.

At $h_0 = 30 \, a_0$, after crossing the first groove, the CW motion reorients the bacterium before it can cross the ridge, and thereby it swims back into the first groove. Similarly to the case $h_0= 20\, a_0$, it swims predominantly along the groove direction. However, in this case, it does not remain aligned with the
$y$-direction.
Instead, a CCW motion causes a large excursion towards the ridge followed by CW swimming along the ``planar part'' of the undulating surface,
which again drives the bacterium back towards the groove.
This pattern repeats several times and it also occurs for $h_0=50 \, a_0$, after the bacterium crosses the ridge
(see Video 3 in the supplementary material).

\begin{figure}
    \centering
    \includegraphics[width=1\columnwidth]{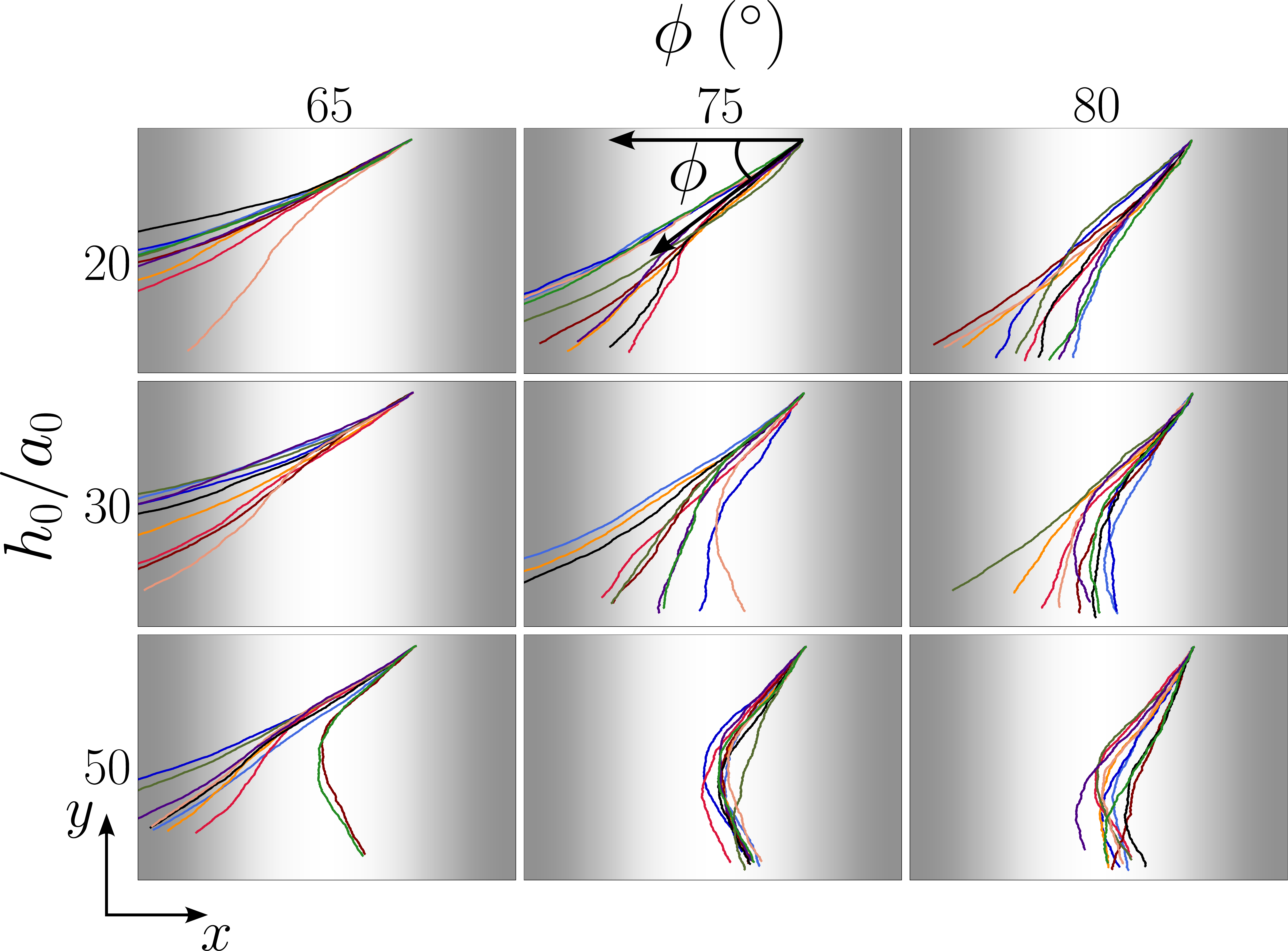}
    \caption{
    Simulated trajectories of individual bacteria attempting to cross the groove for different initial lateral angles $\phi$ measured against the negative $x$ direction
    and amplitudes $h_0$. Ten bacteria are simulated for each pair ($h_0$, $\phi$).
    The gray shading represents the undulating surface, with white
    and dark gray indicating, respectively, the groove and ridge.
    Note that the scales of the $x$ and $y$ axis are different. Therefore, the angles $\phi$ of the trajectories appear smaller than indicated.
}
    \label{fig:cross_groove_attempt}
\end{figure}

These observations suggest that, rather than facilitating escape, large values of $h_0$ enhance bacterial trapping by locking the bacterium into an oscillatory swimming pattern along the groove. To better understand the geometric conditions leading to 
this trapping, we performed simulations in which bacteria approach the groove with varying lateral angles $\phi$ and undulation amplitudes $h_0$. Similarly to Sec.~\ref{subsec.surface_escape}, the bacterium is initialized at $x=0$ with $\alpha=0$,
but now it swims downhill toward the groove.
We report ten trajectories for each pair of parameters
($h_0$, $\phi$) in Fig.~\ref{fig:cross_groove_attempt}.

At $h_0 = 20 \, a_0$, all bacteria successfully cross the groove. Nevertheless, the trajectories become more dispersed as $\phi$ increases,
and at $\phi = 80^\circ$ some trajectories clearly align with the groove.
For $h_0 = 30 \, a_0$, some of the bacteria align with
the groove 
already 
at $\phi=75^\circ$, and, for $\phi=80^\circ$, most bacteria fail to 
cross the groove. 
Instead, they
exhibit CCW motion consistent with the behavior observed in Fig.~\ref{fig:ex_traj_oscillation}. 
At $h_0 =50 \, a_0$, this effect is even more pronounced. 
At $\phi=75^\circ$, none of the bacteria cross the groove, but
swim on CCW trajectories and thereby leave
the groove towards the direction they came from.

Overall, these simulations show that as the undulation amplitude $h_0$ increases, the 
lateral angle $\phi$ must become progressively smaller for the bacterium to successfully cross the groove. 
Thus, for $h_0 \geq 20 \, a_0$, we observe bacterial trapping
since the oscillatory trajectories confine the bacterium 
to the groove.
In Sec.~\ref{subsec.surface_escape}, we found that an increasing
$h_0$ systematically promotes escape, as the
CW motion ultimately brings 
the bacterium to a ridge. Taken together, these results suggest that the surface residence time 
for an undulating surface is not monotonic in $h_0$.
It first decreases with increasing $h_0$ and then increases again.

\subsubsection{Minimal model: rotating rod in a groove}
\begin{figure}
    \centering
    \includegraphics[width=1\columnwidth]{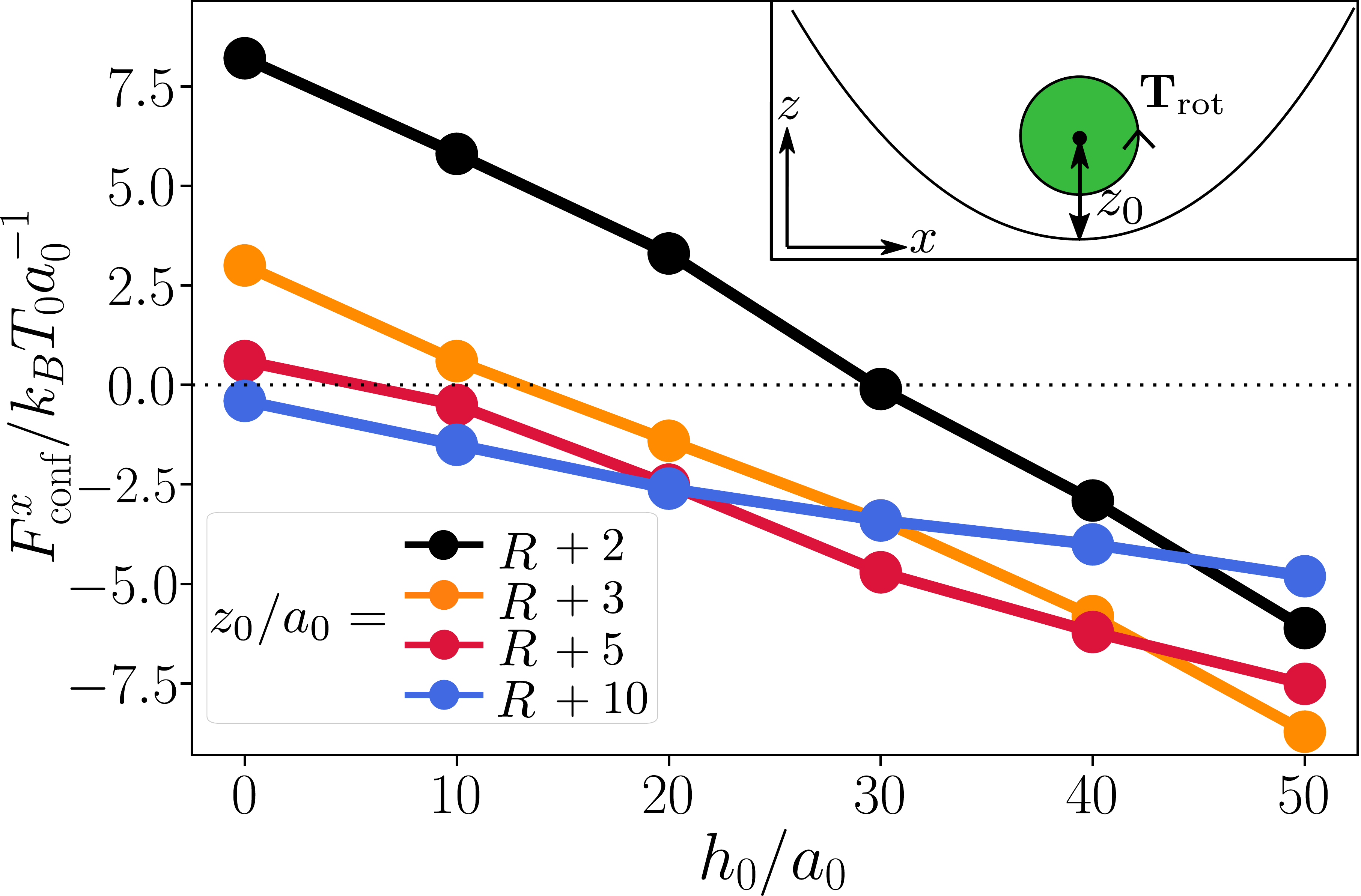}
    \caption{
    Simulated
        confinement force 
        $F^x_{\mathrm{conf}}$ for a rotating rod aligned with the $y$-direction
        plotted
          as a function of the 
          surface-undulation
          amplitude $h_0$,
          for
          different distances $z_0$ from the bottom of the groove. The inset illustrates the simulation set-up, including the definition of the distance $z_0$, and the applied torque.
          The orientation vector $\mathbf{u}_3$ points out of the plane.
}
    \label{fig:rod_exp}
\end{figure}

A transition from CW to CCW motion is typically observed at
planar interfaces to air when the fluid boundary condition switches from no-slip (solid interface) to slip (fluid interface).
\cite{Hu2015PhysicalSlip, Lemelle2010CounterclockwiseInterface} 
In contrast, in
our work, we observe such
a CW to CCW transition inside the grooves,
which is driven by curvature. This suggests
that 
hydrodynamic interactions with a no-slip surface, which are responsible for CW trajectories at planar surface, \cite{Lauga2006SwimmingBoundaries} 
are crucially modified if surfaces become curved.
An obvious candidate to explain this transition is the hydrodynamic
rotlet dipole generated by the rotation of the flagellar bundle and the counter rotation of 
the cell body that interacts with the surface.

To support this hypothesis, we performed simulations of an even simpler situation, namely a
rotating rod inside a groove, with dimensions matching those of the cell body, 
in particular, the rod radius is $R=4.4a_0$.
The rod center is initialized at position $\mathbf{r}_0=\{3\lambda/4, 0, z_0\}$, directly in the groove a height $z_0$, which
we vary in the simulations.
The rod is aligned 
along the groove
($\phi=90^\circ$) and,
therefore, located symmetrically between the rising surfaces on the left and right
[see inset in Fig.~\ref{fig:rod_exp}]. A harmonic trap is used to constrain the position of the rod
with
a confinement force $\mathbf{F}_{\mathrm{conf}}= -K(\mathbf{r}- \mathbf{r}_0)/a_0^2$.
Similarly, we apply a confinement torque
$\mathbf{T}_{\mathrm{conf}}=-K(\mathbf{u}_3\times\mathbf{e}_y)$
to keep the rod axis $\mathbf{u}_3$ aligned with
the $y$-direction
and
set $K=1\times10^5\,k_B T_0$. 
Most importantly,
we apply a torque $\vert\mathbf{T}_{\mathrm{rot}}\vert = 3000\,k_B T_0$
along $\mathbf{u}_3$, so that
the rod 
rotates
around its main axis.
Note that the torque roughly agrees with the total torque generated by the four flagella of our model bacterium.

We perform simulations at different distances from the surface, $z_0$, and for different 
surface-undulation amplitudes, $h_0$. Since the rotation of a rod near a no-slip surface induces a rolling motion, the trapping force measures the impact of  hydrodynamic interactions generated by the nearby surface. We therefore focus on the $x$ component of $\mathbf{F}_{\mathrm{conf}}$, which is reported in Fig.~\ref{fig:rod_exp}. Note that 
using
a confinement force necessarily generates a Stokeslet contribution around the rod, which may alter the measured response. Nevertheless, our aim here is to capture the phenomenology of the mechanism.

For $h_0=0$, \emph{i.e.}, a flat surface, the rotation of the rod induces a translation in the
negative
$x$-direction, 
which generates a confinement force pointing into the positive $x$ direction to stop the translational motion. 
In Fig.\ \ref{fig:rod_exp} this corresponds to the positive values of $F^x_{\mathrm{conf}}$ for small distances $z_0$. However,
for $z_0 = R+5 \, a_0$ and $R+10 \, a_0$ the confinement force along the $x$ direction is nearly zero, although the rod, when turning off the 
harmonic trap, clearly rolls into the negative direction but it is also lifted up. Therefore, we suspect that the Stokeslet flow generated by 
the total confinement force ultimately keeps the rod from translating along the $x$ direction.

For clarity, we concentrate on the distance $z_0 = R+2a_0$, which roughly corresponds to the swimming distance of the model 
bacterium from the undulating surface [see Fig.~\ref{alpha_delta_z}(b)]. Increasing
the surface-undulation amplitude 
decreases
the confinement force towards zero and,
eventually, it becomes negative. Thus, due to the increasing confinement by the undulating surface, stresses are generated that counteract the
viscous stresses from the rolling rod and push the rod in the positive $x$ direction. Now, the model bacterium consists of a rotating
cell body and a counter-rotating bundle of flagella. They roll in opposite directions and thereby the whole bacterium turns in CW direction for
small surface undulations and in CCW direction for large $h_0$, which qualitatively explains our observations.


To further justify our interpretation, we present a second minimal model that relies on analytic calculations.

\subsubsection{Minimal model: rotating rod in a cylindrical cavity}
\label{sec:rollingCylinder}

\begin{figure}
	\centering
   \includegraphics[width=0.90\columnwidth]{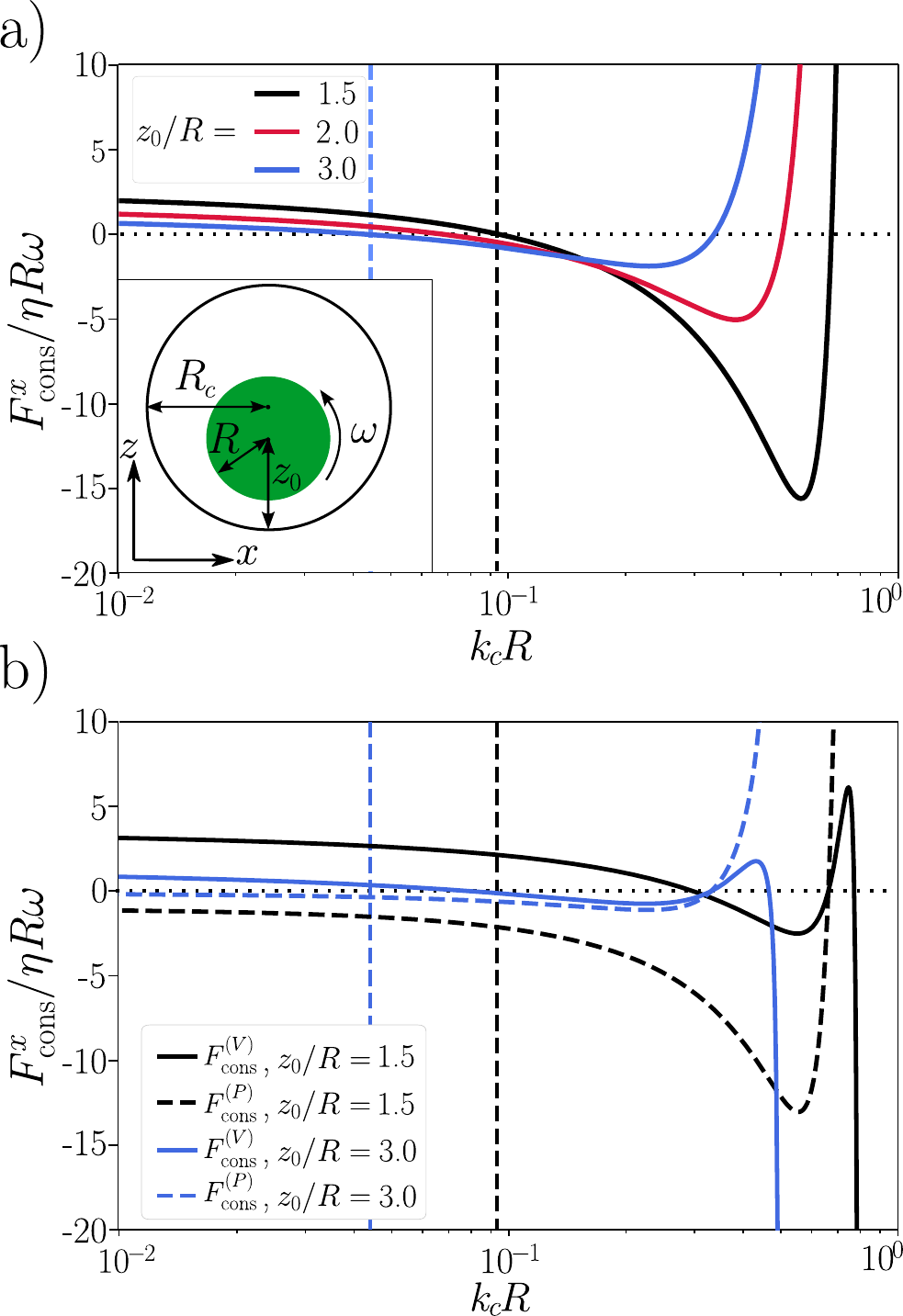}
   \caption{Constraint forces per unit length for the rotating cylinder plotted \emph{vs} cavity curvature $k_c$ for several cylinder heights $z_0$.
    (a) Total constraint force $F^x_\text{cons}$. Vertical dashed lines mark values of $k_cR$ for which $F^x_\text{cons}=0$.
          Inset: Cylinder of radius $R$ rotates with angular frequency $\omega$ in a cylindrical cavity of radius $R_c$. The centers of the cavity
          and cylinder are at $x=0$, and the center of the cylinder is at height $z_0$ above the bottol cavity wall.
    (b) Contributions to the total constraint force, $F^x_\text{cons} = F^{(P)}_\text{cons} + F^{(V)}_\text{cons}$, due to fluid pressure and viscous
          forces. 
}
    	\label{fig:analytics}
\end{figure}

Here, we consider an infinitely long cylinder of radius $R$ that rotates about its long axis with fixed angular speed $\omega$ inside a 
cylindrical cavity, with cylinder and cavity axis parallel to each other [see inset of Fig.\ \ref{fig:analytics}(a)]. The cavity has a radius 
$R_c = 1 / k_c$, where the  curvature $k_c$ is analogous to the undulation amplitude $h_0$. The midpoints of the cavity and the cylinder 
are at $x=0$, while we fix the cylinder's position at a height $z_0$ above the bottom cavity wall [see inset of Fig.\ \ref{fig:analytics}(a)].
For this setup, we solve the Stokes equations using lubrication theory\ \cite{Cope1949,Antunes2022,Antunes2023,Antunes2024, Malgaretti2014,Richter2025, Richter2025_2,Thiele2004,Diekmann2025} as detailed in Appendix \ref{sec:SM_rollingCylinder}.
Then, by integrating the 
Newtonian stress tensor over the cylinder surface,  we determine the hydrodynamic force per unit length, $\mathbf{F} = -\mathbf{F}_\text{cons}$, 
exerted by the fluid on the cylinder. It needs to be balanced by the constraint force $\mathbf{F}_\text{cons}$, in full analogy to $\mathbf{F}_\text{conf}$ 
in the previous subsection. 

In Fig.\ \ref{fig:analytics}(a) we plot the $x$ component $F^x_\text{cons}$ \emph{versus} cavity curvature $k_cR$ for different heights $z_0$.
For small curvature, the positive constraint force keeps the rod from rolling up the cavity wall to the left. However, with increasing $k_c$, the force
tends to zero and becomes negative, in full agreement with our previous results. Since the cylindrical cavity also has an 
upper wall, the constraint force becomes zero again at 
$k_c R = R/z_0$, where cylinder and cavity are concentric, and then diverges at 
$k_cR=2R/(R+z_0)$, where the cylinder touches the upper cavity wall. Of course, this behavior is not seen for the undulating surface.

To shed further light on the sign reversal of the constraint force along the $x$ direction, we separate it into pressure ($F^{(P)}_\text{cons}$) 
and viscous ($F^{(V)}_\text{cons}$) contributions that originate from the respective parts of the stress tensor. In 
Fig.\ \ref{fig:analytics}(b)] we plot both contributions \emph{versus} the cavity curvature $k_cR$ for different heights $z_0$. We concentrate 
on small curvatures, where the cylinder is mostly influenced by the bottom of the cavity, and leave a more thorough discussion to
Appendix\ \ref{sec:SM_rollingCylinder}.
First, the positive viscous contribution to the constraint force dominates.
It resists the cylinder rolling to the left due to the shear flow between cylinder and cavity. However, this shear flow is also driven by a 
nonuniform pressure,
which is higher upstream (to the left of the bottleneck between cylinder and wall) compared to downstream. 
Thus, the fluid pressure pushes the cylinder to the right, 
which is resisted by $F^{(P)}_\text{cons} < 0$. For increasing curvature, the pressure contribution becomes more dominant due to the 
increasing lateral confinement imposed on the cylinder, while $F^{(V)}_\text{cons}$ even decreases, and the total constraint force changes sign.
In total, we can conclude that for larger lateral confinement the fluid pressure is responsible for pushing the cylinder against its rolling direction, 
which also explains the behavior of the rotating rod in the previous subsection.

\section{Summary and conclusions}
\label{sec.conclusion}

In this article, we investigated the motility of a non-tumbling \textit{E.\ coli} strain near corrugated surfaces 
described by sinusoidal height variations.
We employed a fine-scale numerical model that captures the mechanical properties of \textit{E.\ coli}, in which the flexible filaments are 
modeled using the Kirchhoff rod theory. The modeled bacterium swims in an MPCD fluid, which resolves the hydrodynamic flow field 
generated by the bacterium.

We first focused on small surface
undulations, for which persistent surface swimming and CW trajectories are
observed. Such a behavior is expected since, at small curvature, hydrodynamic attraction to
and alignment along the surface
dominate, si\-mi\-lar to experiments of \textit{E.\ coli} swimming
near large cylindrical pillars.\cite{Sipos2015HydrodynamicWalls, Takaha2023Quasi-two-dimensionalDependence} 
For surface-undulation
amplitudes larger than $7\,a_0$, we observe
an attachment-detachment pattern, suggesting that hydrodynamic interactions become insufficient to keep the bacterium close to the surface. Nevertheless, as long as the bacterium remains
near the surface, the typical CW trajectory observed near no-slip surfaces is
recovered.

In a second step, we investigated the escape capability from the ridges of the undulating surfaces
for different bacterial orientations with respect to the undulation direction and for different undulation
amplitudes. We found that the escape rate follows lines of constant curvature, indicating that the surface curvature at the ridge determines bacterial escape. Furthermore, we give an estimate for
a critical curvature $\kappa_0 = 3.4\times10^{-3}\,a_0^{-1}$, above which the majority of bacteria are expected to leave the surface. 

Finally, we performed simulations for undulation amplitudes $h_0 \geq 20\,a_0$, for which we expect most of the bacteria
to escape at the ridge. Surprisingly, we find
that the bacteria persistently swim along the groove direction, with the trajectories oscillating away and toward
the groove. Furthermore, we performed simulations of bacteria initially placed
on the planar or steepest
part of the undulating surface and swimming toward the groove with varying
in-plane angles and surface amplitudes. 
Ultimately, the bacteria become unable to cross the groove, instead, they 
oscillate between CW and CCW swimming and thereby move effectively along the groove. This
results in fewer escape opportunities at the ridge. 
Indeed, the CCW swimming is induced by hydrodynamic interactions of the bacterium with the nearby
confining surface walls at large undulation amplitudes.
To show this, we analyzed the motion
of a rotating cylinder trapped inside the groove. Interestingly, we observe a sign change of the confinement
force at larger undulations
amplitudes $h_0$, which means that the rotating rod drifts into the opposite direction compared to a planar
no-slip surface.
Applying this result to the counter rotating cell body and flagellar bundle, one obtains the CCW motion inside the groove, which turns into a CW motion outside the groove. This explains the observed oscillations. In Ref.~\citenum{Kurzthaler2021MicroswimmersSurfaces}, the sign reversal of the drift force was already anticipated using far-field hydrodynamics, while we show this effect
with a detailed model of \textit{E.\ coli}.
We also performed analytic calculations for a rotating cylinder inside a cylindrical cavity
using lubrication theory. In particular, we find that an increasing pressure force is responsible for the sign change in the drift motion.

In Ref.~\citenum{Perez-Estay2024AccumulationCurvature}, experimental work with
corrugated surfaces revealed bacterial accumulation inside grooves for curvatures larger than $\kappa \geq 0.25\,\mathrm{\mu m}^{-1}$. The authors attributed this result to steric interactions. 
In contrast, in our article we demonstrate that a pure hydrodynamic effect can initiate bacterial trapping already at smaller curvatures.
However, the extent to which this mechanism contributes to bacterial accumulation and potential biofilm formation
requires further investigations. Nevertheless, our results demonstrate that hydrodynamic effects influence the motion of \emph{E.\ coli} also near complex surfaces.

The versatility of our approach allows for further investigations of \textit{E.\ coli} motility near surfaces. In this work, we focused on a non-tumbling strain. How tumbling is modified near corrugated or more confined spaces remains to be determined. Furthermore, bacteria in their biological environments are often exposed to flows. Investigating their motility in sheared flows near corrugated surfaces could therefore provide a more realistic description of bacterial transport near complex topographies.



\section*{Conflicts of interest}
There are no conflicts to declare.

\section*{Data availability}
The data supporting the findings of this study are available upon request from one of the authors.

\appendix

\section{Lubrication theory for a confined rolling cylinder}

\label{sec:SM_rollingCylinder}

\begin{figure}
	\centering
   \includegraphics[width=0.9\columnwidth]{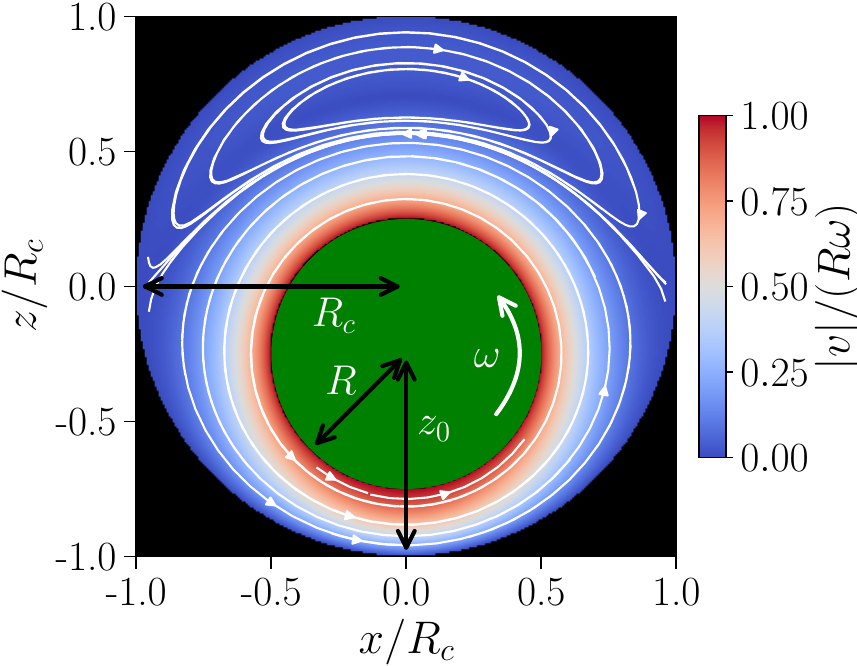}
   \caption{Snapshot of flow field for $k_cR = 1/2, z_0/R=1.5$. For these parameters, the constraint force on the cylinder points to the left.  
   The fluid streamlines are shown in white and the red/white/blue spectrum indicates the magnitude of the velocity. The rolling cylinder is 
   portrayed in green and the static cavity wall in black. Lengths such as the respective 
radii of the cylinder and cavity, $R$ and $R_c$, the height $z_0$ of the cylinder above the bottom cavity wall, and the angular speed $\omega$ of the cylinder are portrayed.
}
    	\label{fig:snapshot}
\end{figure}

We now derive the lubrication theory used to produce the results in Section \ref{sec:rollingCylinder}. We consider an infinitely long cylinder of radius $R$ rolling along its long axis with angular speed $\omega$ [see inset of Fig.\ \ref{fig:analytics}(a)]. 
This cylinder is inside an infinitely long cylindrical cavity of radius $R_c$ oriented parallel to the cylinder's axis of symmetry. We define the curvature of the cavity as $k_c=1/R_c$. Between the cylinder and the cavity walls is a Newtonian fluid of viscosity $\eta$. The cylinder 
rotates with an angular velocity $\omega$ and the hydrodynamic force $\mathbf{F} = -\mathbf{F}_\text{cons}$ exerted by the fluid 
(per unit length of the cylinder) on the cylinder is counterbalanced by the external constraint force $\mathbf{F}_\text{cons}$.

The fluid is ruled by the Stokes equation together with the incompressibility condition,
\begin{equation}
    \eta \Delta \bm{v}(\bm{r}) =  \nabla P(\bm{r}) \enspace \mathrm{and} \enspace   \nabla \cdot \bm{v}(\bm{r}) = 0 \, , 
    \label{eq_SM_incompressibility}
 \end{equation}
with $\bm{v}(\bm{r})$ being the fluid velocity and $P(\bm{r})$ 
the pressure. In the frame of reference centered on the cavity, the axis of symmetry of the cylinder is located at $(x,z)=(0, -d)$. Now, it is convenient to use polar coordinates in the $x,z$ plane and centered on the cylinder. 
The boundaries of the fluid are thus represented by $r=R$
for the surface of the cylinder and
\begin{equation}
    r = R' = d \cos(\phi - \pi/2) + \sqrt{ R_c^2 - d^2\sin^2(\phi-\pi/2)},
    \label{eq_cavitywall}
\end{equation}
%
%
for the surface of the cavity wall, where $R^{\prime}(\phi)$ is the location of the cavity wall. 
The boundary conditions are 
\begin{equation}
    v_{\phi} (r = R) = R \omega \, , \enspace v_{r} (r = R) = 0 \, , \enspace \mathrm{and} \enspace   \bm{v} [r = R^{\prime}(\phi)] = 0 \, ,
    \end{equation}
where $v_{\phi}(r,\phi)$ is the azimuthal component of the velocity and $v_r(r,\phi)$ is the radial component of the velocity. Note that we have used the homogeneity of the velocity field along $y$. 

The Stokes equation for the azimuthal component yields
\begin{equation}
    \frac{1}{r} \frac{\partial_\phi P(\phi)}{\eta} = - \frac{v_\phi(r,\phi)}{r^2} + \frac{2}{r^2} \partial_{\phi}v_r(r,\phi) + \frac{1}{r}\partial_r(r \partial_r v_{\phi}) + \frac{1}{r^2}\partial_{\phi}^2 v_{\phi} \label{stokes_beforeLub}
\end{equation}
We now estimate the order of magnitude of each term in order to apply the lubrication approximation. First, we outline the principal idea and then we are more specific. The order of magnitude of the derivative of a function is the order of magnitude of variations of the function 
divided by the order of magnitude of the typical variation of the argument, such that
\begin{equation}
    \partial_r v_{\phi}\approx \frac{ \overline{v}_{\phi}}{\Delta R}, \label{eq_lub}
\end{equation}
where $\overline{v}_{\phi}$ is the typical azimuthal speed and  $\Delta R$ is the typical thickness of the film between the cavity and the cylinder. Approximating the derivatives with respect to $r$ in Eq. \eqref{stokes_beforeLub} using the formula in Eq. \eqref{eq_lub}, we obtain
\begin{equation}
    \frac{1}{r} \frac{\partial_\phi P(\phi)}{\eta} =   \frac{1}{r}\partial_r(r \partial_r v_{\phi}(r,\phi)) + \mathcal{O}\left[ \left( \frac{\Delta R}{\overline{R}} \right)^2 \right],
\end{equation}
where $\overline{R}$ is the typical value of $r$, and where we have also applied the same procedure to the incompressibility condition 
in 
Eq. \eqref{eq_SM_incompressibility},
\begin{equation}
    \partial_{\phi}v_{\phi}(r,\phi) = - \partial_r[r v_r(r,\phi)],
\end{equation}
yielding
\begin{equation}
    v_r(r,\phi) \approx \frac{ \Delta R}{\overline{R}} v_{\phi} (r,\phi).
\end{equation}
Taking the lubrication approximation,
$ \Delta R / \overline{R} \ll 1$,
we approximate Eq. \eqref{stokes_beforeLub} as
\begin{equation}
    \frac{1}{r} \frac{\partial_\phi P(\phi)}{\eta} =   \frac{1}{r}\partial_r(r \partial_r v_{\phi}(r,\phi)) \, . \label{stokes_afterLub}
\end{equation}
The range of parameters for which 
$ \Delta R / \overline{R} \ll 1$
is valid can be estimated by the following calculation: the quantity $\Delta R$ can be approximated by the average of its highest and lowest values ($R_c +d - R$ and  $R_c -d - R$, respectively), yielding $\Delta R \approx R_c -R$. The quantity $\overline{R}$ may be likewise estimated as the average of its highest ($2R_c -R$) and lowest ($R$) value, yielding $\overline{R} \approx R_c$
As such, we estimate
\begin{equation}
    \frac{\Delta R}{\overline{R}} \approx 1-k_cR
\label{eq.estimate}
\end{equation}
which only takes values lower than one. As such, the lubrication approximation is expected to yield results that are at least qualitatively 
correct for a wide region of parameter space. 

We may now integrate Eq. \eqref{stokes_afterLub} along $r$ twice to yield the velocity field
\begin{equation}
    v_{\phi}(r,\phi) = f_1(r,\phi) \partial_{\phi} P(\phi) + f_2(r,\phi) \omega,
    \label{velocityfield}
\end{equation}
where 
\begin{equation}
    f_1(r,\phi) = -\frac{1}{\eta} \left[ R' -r + \frac{\log [ r/R' ]}{\log[R/R']} [R - R'] \right],
\end{equation}
and
\begin{equation}
    f_2(r,\phi) = \frac{\log [ r/R' ]}{\log[R/R']} R.
\end{equation}
These two auxiliary functions can be computed as they depend only on the known parameters of the system. 

It can also be shown that the incompressibility condition in Eq. \eqref{eq_SM_incompressibility} requires the volumetric flux $Q$ through a line running from the center of the cylinder to the cavity wall,
\begin{equation}
    Q = \int\limits_{R}^{R'} v_{\phi}(r,\phi) dr  \, ,
\end{equation}
to be independent of $\phi$. Using the velocity field in Eq. \eqref{velocityfield}, we obtain 
\begin{equation}
    Q = f_3(\phi) \partial_{\phi} P(\phi) + f_4(\phi) \omega,
    \label{Q}
\end{equation}
with
\begin{equation}
    f_3(\phi) = \frac{R - R'}{2 \eta  \log[R/R']}\{-2R + 2R' + [R+R'] \log[R/R'] \},
\end{equation}
and
\begin{equation}
    f_4(\phi) = \frac{R}{\log[R/R']} \{ R - R' + R\log[R'/R] \}.
\end{equation}
Equation \eqref{Q} can also be written as 
\begin{equation}
    \partial_{\phi} P(\phi) = \frac{Q - f_4(\phi) \omega}{f_3(\phi)},
\end{equation}
which we integrate over $\phi$ to obtain
\begin{equation}
    \int\limits_0^{2\pi} \partial_{\phi} P(\phi) d\phi = 0 = Q \INTPHI f_3^{-1}(\phi) - \omega \INTPHI f_4(\phi) f_3^{-1}(\phi), 
\end{equation}
or
\begin{equation}
    Q = \omega \left[ \INTPHI f_3^{-1}(\phi)\right]^{-1}\INTPHI f_4(\phi) f_3^{-1}(\phi), 
\end{equation}
which gives a closed-form expression for the pressure
\begin{equation}
    \partial_{\phi} P(\phi) = \omega \left[ \frac{\INTPHI f_4(\phi) f_3^{-1}(\phi)}{f_3(\phi) \INTPHI f_3^{-1}(\phi)}  - \frac{f_4(\phi)}{f_3(\phi)}\right]. \label{eq_SMIntegratePressure}
\end{equation}
If the cylinder is concentric with the cavity ($d=0$), then $R^{\prime}$ is independent of $\phi$, and so is $f_3$ and $f_4$. As such, the pressure 
is homogeneous, no matter how fast the cylinder spins. 

The force density $\bm{f}(\phi)$ on the surface of the cylinder is given by 
\begin{equation}
    \bm{f}(\phi) = \bm{\sigma} (\phi, r=R) \cdot \bm{n}(\phi),
\end{equation}
where $\bm{n}(\phi)=  \bm{e}_r(\phi)$ is the radial unit vector,
normal to the cylinder surface, and
$\bm{\sigma}$ is the stress tensor
\begin{equation}
    \bm{\sigma}(r,\phi) =-P(\phi) \bm{I} + \eta [\nabla \bm{v}(r,\phi) + \nabla \bm{v}^T(r,\phi)].
\end{equation}
In cylindrical coordinates, the force density is written as
\begin{align}
    \bm{f}(\phi,r)  = &[ 2\eta\partial_r v_r(r,\phi) - P(\phi)] \bm{e}_r(\phi) + \nonumber \\
     &+ \eta\left[ \partial_r v_{\phi}(r,\phi) + \frac{1}{r} \partial_{\phi}v_r(r,\phi) - \frac{v_{\phi}(r,\phi)}{r}\right] \bm{e}_{\phi}(\phi), \label{eq_forcedensity}
\end{align}
where the last factor on the right-hand-side is the azimuthal unit vector. The total force exerted by the fluid on the cylinder (per unit length of the cylinder) is then the integral
\begin{equation}
    \mathbf{F} = \INTPHI \bm{f}(\phi,r = R). \label{eq_SMForce}
\end{equation}

Now, we calculate $\mathbf{F} $ explicitly. From the boundary conditions it
follows that $\partial_{\phi} v_r(\phi, r =R) =0$,
and the incompressibility condition (Eq. \eqref{eq_SM_incompressibility}) yields
\begin{equation}
    \partial_r v_r(\phi, r =R) = - \frac{1}{R} \partial_{\phi} v_{\phi} (\phi, r= R) - \frac{v_r(\phi,r=R)}{R} \, .
\end{equation}
Since both terms on the right-hand-side are zero, we obtain $\partial_r v_r(\phi, r =R) = 0$. The above results can be plugged into Eq. \eqref{eq_SMForce} yielding
\begin{align}
    \frac{\mathbf{F}}{\eta} = \INTPHI &\left\{ R \left[ \frac{\partial_{\phi}P(\phi)}{\eta} \frac{R' - R + R\log[R/R']}{R \log[R/R']} + \right. \right. \nonumber \\
    &\left. \left. + \frac{\omega}{\log[R/R']} - \omega\right]\bm{e}_{\phi}(\phi) - \frac{P(\phi)}{\eta} \bm{e}_r(\phi)\right\}.
\end{align}
The $x$ component of this force is
\begin{align}
    \frac{F_x}{\eta} = \INTPHI &\left\{ -R \left[ \frac{\partial_{\phi}P(\phi)}{\eta} \frac{R' - R + R\log[R/R']}{R \log[R/R']} + \right. \right. \nonumber \\
    & \left. \left.+\frac{\omega}{\log[R/R']} - \omega\right]\sin(\phi) - \frac{P(\phi)}{\eta} \cos(\phi)\right\}, \label{eq_SMForcex}
\end{align}
where the pressure must be obtained by integrating $\partial_{\phi} P(\phi)$ as per Eq. \eqref{eq_SMIntegratePressure}, which in the general case must be done numerically. Note that the pressure is defined up to an additive constant, but that has no effect on the force. It can be seen that the quantity $F_x / \eta$ is independent of $\eta$ and linear in $\omega$. 

The force can be split into a pressure contribution $F_x^{(P)}$ coming from the pressure term of the stress tensor and a viscous contribution $F_x^{(V)}$ coming from the viscous term of the stress tensor,
\begin{equation}
    \frac{F_x^{(P)}}{\eta} = \INTPHI \left[  - \frac{P(\phi)}{\eta} \cos(\phi)\right], \label{eq_SMForcexP}
\end{equation}
and
\begin{align}
    \frac{F_x^{(V)}}{\eta} = \INTPHI &\left\{ -R \left[ \frac{\partial_{\phi}P(\phi)}{\eta} \frac{R' - R + R\log[R/R']}{R \log[R/R']} + \nonumber \right.\right. \\
    & \left. \left. + \frac{\omega}{\log[R/R']} - \omega\right]\sin(\phi)\right\}. \label{eq_SMForcexV}
\end{align}
Numerically integrating Eqs. \eqref{eq_SMIntegratePressure} and \eqref{eq_SMForcex} - \eqref{eq_SMForcexV} together with the 
condition $\mathbf{F} = -\mathbf{F}_\text{cons}$ results in the data shown in the main text (Fig. \ref{fig:rod_exp}).

The main features of this theory are discussed in the main text and Fig.\ \ref{fig:analytics}.
In addition, we observe a region 
for $k_cR$
where the pressure and viscous contributions point in the same (positive) 
direction
or where the constraint forces in Fig.\ \ref{fig:analytics}(b) 
both become negative.
This phenomenon  may be due to the formation of a vortex on top of the cylinder that rotates in the opposite direction of the cylinder (see Fig. \ref{fig:snapshot}). 
As such, the layer of fluid which is co-rotating with the cylinder does not show a large difference in thickness when comparing the region 
below and above the cylinder. Thus, the viscous drag on the upper half of the cylinder (pointing to the right) is comparable with the viscous drag on the lower half of the cylinder (pointing to the left).

When $k_cR = R/z_0$, the cylinder and the cavity are concentric. Due to radial symmetry, the hydrodynamic force acting on the cylinder is 
zero, as can be seen in Fig. \ref{fig:analytics}. It is also readily readable from Eq. \eqref{eq_SMForcex} as then $R^{\prime}$ and $P$ 
take homogeneous values. For yet higher curvatures $k_cR > R/z_0$  the hydrodynamic force (and the two components) flip sign. This 
observation occurs due to the bottleneck between cylinder and cavity wall being now located above the cylinder (rather than below). The 
maximum possible curvature is $k_cR=2R/(R+z_0)$, at which point the upper cavity wall touches the cylinder and the hydrodynamic force diverges.

\section*{Acknowledgements}
We thank Tapan Adhyapak, Arne Zantop, and Zihan Tan for useful discussions on the development  of the model 
as well as Benjamin P\'erez-Estay and Anke Lindner for helpful insights on the behavior of \textit{E.\ coli} close to corrugated surfaces.
Financial support from the Deutsche Forschungsgemeinschaft (DFG) (Grant number 462445093) and TU Berlin
is gratefully acknowledged.

\bibliography{ref} 
\bibliographystyle{rsc} 
\end{document}